\documentclass[9pt,twocolumn]{article}
\usepackage[margin=1in]{geometry}
\usepackage{graphicx}
\usepackage{subcaption}
\usepackage{booktabs}
\usepackage{caption}
\usepackage{multicol}
\usepackage{tabularx}
\usepackage[numbers]{natbib}
\usepackage{hyperref}
\usepackage{amsmath}
\usepackage{mathptmx}

\title{QDockBank: A Dataset for Ligand Docking on Protein Fragments Predicted on Utility-Level Quantum Computers}

\author{
Yuqi Zhang$^{1}$, Yuxin Yang$^{2}$, Cheng-Chang Lu$^{3}$, Weiwen Jiang$^{4}$, \\
Feixiong Cheng$^{2}$, Bo Fang$^{5}$, Qiang Guan$^{1,*}$ \\
\\
$^1$Kent State University, Kent, OH, USA \\
$^2$Cleveland Clinic, Cleveland, OH, USA \\
$^3$Qradle Inc, Kent, OH, USA \\
$^4$George Mason University, Fairfax, VA, USA \\
$^5$Pacific Northwest National Laboratory, Richland, WA, USA \\
\texttt{qguan@kent.edu}
}
\date{}

\begin{document}


\maketitle


%

%

%



\begin{abstract}
Protein structure prediction is a core challenge in computational biology, particularly for fragments within ligand-binding regions, where accurate modeling is still difficult. Quantum computing offers a novel first-principles modeling paradigm, but its application is currently limited by hardware constraints, high computational cost, and the lack of a standardized benchmarking dataset. In this work, we present \textit{QDockBank}—the first large-scale protein fragment structure dataset generated entirely using utility-level quantum computers, specifically designed for protein–ligand docking tasks. QDockBank comprises 55 protein fragments extracted from ligand-binding pockets. The dataset was generated through tens of hours of execution on superconducting quantum processors, making it the first quantum-based protein structure dataset with a total computational cost exceeding one million USD. Experimental evaluations demonstrate that structures predicted by QDockBank outperform those predicted by AlphaFold2 and AlphaFold3 in terms of both RMSD and docking affinity scores. QDockBank serves as a new benchmark for evaluating quantum-based protein structure prediction.
\end{abstract}







\section{Introduction}

With the rapid development of quantum computing, computational biology is undergoing a paradigm shift, particularly in the modeling and simulation of complex biomolecular systems, where quantum methods are beginning to demonstrate substantial potential. Among these challenges, protein structure prediction—especially the modeling of functional regions such as ligand-binding pockets—remains one of the most difficult tasks in the field~\cite{liang1998anatomy}. Traditional computational approaches, such as classical molecular dynamics simulations or empirical algorithms, often suffer from high computational cost and limited accuracy due to the vast conformational landscapes and intricate energy surfaces of proteins. Recent advances in deep learning have significantly enhanced structural prediction accuracy. In particular, the AlphaFold~\cite{abramson2024accurate,jumper2021highly, evans2021protein} series has established an end-to-end modeling framework based on large-scale residue contact maps and attention mechanisms, and its predictive accuracy has surpassed many physics-based methods, making it a leading approach in the field. However, deep learning models rely heavily on large-scale, high-quality training datasets and tend to capture statistical distributions of known structures rather than fundamental physical principles. As a result, they exhibit a strong ``prior bias''. For structurally diverse, short protein segments (e.g., 5–20 residues) commonly found in active pockets, data sparsity and high variability often lead to significant performance degradation. This indicates that existing models still fall short of capturing the underlying principles governing protein conformational evolution, limiting their generalizability and interpretability in biological contexts.

Quantum computing presents a fundamentally different approach, leveraging quantum superposition and entanglement to potentially model complex molecular systems directly from first principles. This capability allows more efficient exploration of rugged energy landscapes, particularly relevant to protein structures~\cite{wang2024efficient}. However, existing studies typically validate only a few small fragments or rely on simulated quantum circuits due to high computational costs and hardware limitations associated with real quantum executions. A major bottleneck is the prohibitive expense of quantum computation: commercial access to IBM quantum processors is billed by the minute, and predicting a single fragment can require hours of runtime to achieve acceptable accuracy. Furthermore, the absence of standardized benchmarks impedes fair and reproducible performance comparisons, hindering further advancements in quantum biomolecular modeling.

To address these issues, we introduce \textit{QDockBank}, the first large-scale protein structure dataset generated exclusively using real quantum devices, specifically designed for protein–ligand docking applications. We propose that a standardized, reusable, and fully quantum-generated dataset is essential for developing and evaluating quantum-assisted modeling algorithms, thereby enhancing reproducibility in this emerging research field. \textit{QDockBank} focuses on ligand-binding pocket fragments extracted from real proteins, suitable for both structure modeling and downstream docking tasks. Notably, our experimental results demonstrate that the quantum-generated protein structures in \textit{QDockBank} outperform AlphaFold3(AF3) in structural accuracy and functional evaluations. Specifically, compared to experimentally determined X-ray structures, predictions in \textit{QDockBank} exhibit lower root-mean-square deviation (RMSD) values and yield more favorable ligand-binding affinity scores in docking simulations. This highlights the unique advantages of quantum-first modeling approaches, especially in scenarios where deep learning models struggle due to insufficient context. \textit{QDockBank} is built on IBM's utility-level superconducting quantum processors and employs the Variational Quantum Eigensolver (VQE) framework for energy minimization. Each entry in \textit{QDockBank} includes the predicted 3D structure, RMSD relative to the experimental crystal structure, ligand-binding affinity score, and comprehensive quantum metadata (including qubit count, circuit depth, and energy distribution). The current version of \textit{QDockBank} contains 55 representative protein fragment sequences of various biological functions and structural classes, all derived from known biologically active ligand-binding regions. These fragments span a range fully compatible with the modeling capabilities of current quantum hardware. The dataset comprises more than 2,000 docking tests and hundreds of thousands of quantum circuit executions, totaling dozens of hours of quantum processor runtime—making it the most extensive and computationally expensive protein structure dataset ever produced on a physical quantum computer to date.

The main contributions of our work are as follows: (1). We introduce the first protein structure prediction dataset generated entirely on real quantum hardware, along with a comprehensive evaluation framework tailored for quantum-based prediction, supporting a wide range of downstream applications. (2). The predictions in \textit{QDockBank} outperform those of current leading deep learning methods, with over 90\% of entries surpassing AlphaFold2 and more than 80\% surpassing AlphaFold3 in accuracy. (3). The dataset was generated using over 60 hours of quantum processor runtime.

The remainder of this paper is structured as follows: Section~2 reviews existing structure databases and quantum-based methods; Section~3 details the design and contents of \textit{QDockBank}; Section~4 explains the quantum multi-tasking processes; Section~5 evaluates structural accuracy and diversity; and Section~6 outlines potential applications. The dataset is freely available at: \url{https://github.com/QDockBank/QDockBank_}.

\section{Background}

\subsection{Protein prediction and docking:} Protein structure prediction is the computational task of determining a protein’s three-dimensional conformation from its amino acid sequence. This task is fundamental in life sciences and drug development, particularly for modeling functional regions such as ligand-binding pockets. \textit{Ligand–receptor docking} simulates how a small molecule (ligand) fits into and interacts with a specific binding site on a protein (receptor)~\cite{shoichet1991protein}. Analogous to a key fitting into a lock, the interaction is driven by non-covalent forces—including hydrogen bonding, hydrophobic interactions, and electrostatic attraction—and may induce local conformational changes that stabilize the ligand–protein complex. These interactions are central to biological processes like signal transduction and enzymatic catalysis, and thus form the basis of rational drug design. For example, targeted cancer therapies aim to inhibit protein active sites using small-molecule inhibitors that block signaling pathways. The primary goal of structure-based docking is to predict the optimal protein conformation upon ligand binding and evaluate interaction quality. Two key evaluation metrics are commonly used: \textit{binding affinity} and \textit{root-mean-square deviation (RMSD)}. Affinity (in kcal/mol) reflects binding free energy, with lower values indicating stronger interactions. RMSD measures geometric deviation from experimental structures and serves as a standard indicator of structural accuracy. Both metrics are widely used in structure prediction, virtual screening, and drug discovery.

\subsection{Quantum computing for protein prediction} Quantum computing has recently emerged as a promising paradigm for modeling molecular systems from first principles. Unlike classical approximations or data-driven deep learning models, quantum methods leverage fundamental physical properties—such as superposition and entanglement—to explore conformational states within high-dimensional Hilbert spaces. Algorithms like the \textit{Variational Quantum Eigensolver (VQE)} are well suited to current utility-level quantum processors, enabling hybrid quantum–classical optimization of protein energy landscapes~\cite{robert2021resource}. While deep learning methods rely heavily on prior data and often struggle with short-chain predictions due to limited informational content, quantum approaches optimize directly from physical laws, rendering their performance largely independent of sequence length. Despite this advantage, practical deployment remains constrained by high hardware costs, system instability, and limited access to quantum resources. As a result, most quantum algorithm validations still depend on classical simulations. Although recent studies have demonstrated the feasibility of coarse-grained protein fragment modeling on real quantum hardware, these efforts remain small in scale, lack standardized benchmarks, and often omit docking validation—hindering comprehensive evaluation and broader adoption~\cite{doga2024perspective}.

\textbf{RCSB PDB Standardized Representation:} In this study, all proteins are referenced by their \textbf{PDB ID}—a unique four-character alphanumeric code assigned by the Protein Data Bank (PDB)~\cite{berman2007worldwide}. Referencing proteins by their PDB IDs ensures traceability, standardization, and reproducibility, allowing readers to directly access the original structural data through public databases. Throughout this paper, we use PDB IDs (e.g., \texttt{4jpy}, \texttt{3d83}) to denote the protein structures from which fragments were extracted for quantum modeling and docking evaluation.

\section{Related Work}

Protein structure prediction has long relied on curated datasets to support model training, benchmarking, and evaluation. Over the past two decades, numerous influential resources have emerged to support computational biology and bioinformatics. One of the most foundational and widely used databases is the \textbf{Protein Data Bank (PDB)}, which provides experimentally determined three-dimensional structures of biological macromolecules~\cite{sussman1998protein, burley2017protein, berman2007worldwide, berman2002protein}. As a cornerstone of structural biology, the PDB has supported classical methods such as molecular dynamics (MD) simulations and comparative modeling. However, due to the time-consuming and resource-intensive nature of experimental methods, the PDB represents only a small fraction of the vast protein sequence space, leaving many biologically relevant structures unresolved. To address this limitation, structural classification databases such as \textbf{CATH}~\cite{orengo1999cath, pearl2003cath} and \textbf{SCOP}~\cite{murzin1995scop} were developed to organize known protein structures into hierarchical families based on evolutionary and structural similarities. Although these resources provide valuable structural annotations and enable comparative analyses, they lack predictive structural data, especially at the fragment level, such as ligand-binding pockets. 

With the advent of deep learning, the demand for large-scale predicted structure datasets has grown considerably. The \textbf{AlphaFold Protein Structure Database}~\cite{varadi2022alphafold, Mihaly2024alphafold}, created by DeepMind and EMBL-EBI, marked a major breakthrough by providing high-confidence structural predictions for over 200 million protein sequences. This resource dramatically expanded structural coverage and rapidly became a foundational reference in computational biology. Nevertheless, AlphaFold predictions are less reliable at fragment-level modeling, particularly for highly flexible, disordered, or short peptide sequences often located in active or ligand-binding regions~\cite{holcomb2023evaluation}. Specialized databases such as \textbf{SwissSidechain}~\cite{gfeller2012swisssidechain} and \textbf{PDBbind}~\cite{wang2005pdbbind, liu2015pdb} have focused on ligand-binding sites and protein–ligand interactions. These resources are indispensable in structure-based drug discovery and docking studies, offering valuable annotations of affinity data and structural complexes. However, these databases primarily rely on classical modeling pipelines and do not incorporate quantum-generated predictions or quantum-derived fragment-level structures. Collectively, these resources underscore the critical role of structured and curated datasets in advancing protein modeling. At the same time, they expose a notable gap: the absence of benchmark-quality datasets generated using quantum algorithms, particularly for short, functional fragments such as ligand-binding pockets.

\textbf{QDockBank} aims to fill this gap by introducing the first large-scale quantum-generated dataset specifically designed for structure-based docking. It complements existing resources by providing fragment-level quantum-predicted protein structures, reproducible docking results, RMSD-based structural accuracy metrics, binding affinity evaluations, and detailed quantum metadata. \textit{QDockBank} thus establishes a critical foundation for benchmarking quantum-assisted protein modeling workflows and facilitates systematic evaluation of quantum computing’s utility in structural biology.

\section{Dataset Design}

\subsection{Data Selection}

The protein fragments in \textit{QDockBank} were selected based on two primary criteria: biological functional relevance and compatibility with current quantum hardware limitations. All fragments in the dataset are derived from real proteins included in the PDBbind dataset, specifically from ligand-binding pockets or their adjacent regions. These regions are essential for molecular recognition, catalytic activity, and protein–ligand interactions; thus, accurate prediction of their local 3D structures is critically important for drug discovery and functional protein studies. From a computational standpoint, our selection strategy carefully considered practical limitations of current utility-level quantum processors, including available qubit count, circuit depth, and coherence time. To ensure each structure could be encoded into a Hamiltonian, optimized, and measured within the capacity of real quantum hardware, we limited fragment lengths to between 5 and 14 residues. To facilitate controlled experimentation, the dataset was divided into three groups based on fragment length. This grouping enables performance evaluation at varying quantum circuit complexities and facilitates the analysis of trade-offs between computational resource usage and prediction accuracy. 

Additionally, we emphasized structural and functional diversity during fragment selection. The 55 fragments originate from distinct PDB entries, covering a wide range of protein functions and structural types. These fragments include proteins derived from animal, plant, and viral sources, encompassing enzymes and other functionally significant proteins. For instance, selected fragments from proteins such as \texttt{1yc4}, \texttt{4jpx}, \texttt{5cqu}, \texttt{3d7z}, \texttt{1ppi}, and \texttt{6p86} represent diverse structural types and biological functions, thereby improving the generalizability of the dataset. Certain fragment sequences, such as \texttt{EDACQGDSGG} and \texttt{LLDTGADDTV}, appear across multiple protein contexts, enabling examination of quantum model generalization across diverse structural environments. Many selected sequences are enriched with highly polar residues (e.g., Asp, Glu, Lys, Arg) and hydrophobic residues (e.g., Leu, Val, Met, Ile), indicative of critical functional regions such as binding interfaces or active sites. Furthermore, several sequences contain conserved structural motifs (e.g., \texttt{GDSGG}, \texttt{LLDTGADDTV}, \texttt{YLVTHLMGAD}), characteristic of enzymatic cores or catalytic signatures.

Importantly, all fragments are associated with experimentally validated X-ray crystal structures and corresponding ligand data, enabling precise assessment of structural accuracy via RMSD and functional relevance through ligand docking affinity scores. Thus, the dataset is not only computationally tractable for quantum modeling but also biologically meaningful and experimentally grounded, making it a representative and valuable resource for quantum-assisted protein structure prediction.

\subsection{Dataset Structure}

Fragments in the dataset were categorized based on their sequence lengths, ranging from 5 to 14 residues. The dataset is organized into three groups:

\begin{itemize}
    \item \textbf{S group:} Fragments containing 5–8 residues
    \item \textbf{M group:} Fragments containing 9–12 residues
    \item \textbf{L group:} Fragments containing 13–14 residues
\end{itemize}

Within each group, multiple subfolders are included, each named after a corresponding protein PDB ID. Each subfolder contains three files corresponding to the following dataset components:

\textbf{Predicted Structure:} This includes the final predicted protein structures in PDB format. These PDB files can be directly visualized or integrated into downstream biological applications, designed to be accessible even for researchers unfamiliar with quantum computing or computational biology.

\textbf{Quantum Prediction Metadata:} Metadata for each prediction is stored in a JSON file, including fragment sequences, number of qubits utilized, and the minimum energy achieved during quantum optimization, facilitating reproducibility and enabling replication by other researchers. Tables~\ref{tab_l}, \ref{tab_m}, and \ref{tab_s} summarize detailed protein information for groups L, M, and S, respectively. Each table lists fragment sequences, their lengths, and the corresponding regions within the complete protein structures. Additionally, each table summarizes key parameters from the Variational Quantum Eigensolver (VQE) optimization process, including the number of qubits required, quantum circuit depth after parameterization, and associated energy metrics—the energy range, minimum, and maximum energies of each quantum system during optimization. The actual execution time required for each fragment prediction is also reported. 

To evaluate computational demands in QDockBank, we analyzed prediction metadata across fragment groups. In terms of qubit count, the L group ranged from 92 to 102 (avg. 98.2), the M group from 54 to 102 (avg. 79.4), and the S group from 12 to 46 (typical value: 23). Circuit depth followed a similar trend: S averaged 127, M 262, and L 396, with several L-group fragments reaching the maximum depth of 413. Energy range was highest in the L group (avg. 6883.6, max. 9200.3 in \texttt{5nkb}), followed by M (2961.7) and S (541.7), indicating increasing conformational complexity. Execution time varied widely: L group ranged from 7,620 s to 199,292 s, M group had an outlier \texttt{4y79} at 207,445 s, while most S-group fragments fell between 4,000 and 20,000 s, with \texttt{5c28} as an exception (114,029 s). These patterns reveal a gradient of computational complexity across groups, with L suited for high-load benchmarking, M for balanced testing, and S for efficient evaluation on low-resource quantum devices.

\begin{table*}[htbp]
\footnotesize
\centering
\caption{L group proteins}
\resizebox{\linewidth}{!}{
\begin{tabular}{lllr|rrrrrr}
\toprule
PDB ID & Residue Sequence & Seq. Length & Residues & Qubits & Depth & Lowest Energy & Highest Energy & Energy Range & Exec. Time (s) \\
\midrule
1yc4  & ELISNSSDALDKI   & 13 & 47--59     & 92  & 373 & 16129.383 & 20745.807 & 4616.425 & 15777.29 \\
3d7z  & YLVTHLMGADLNNI  & 14 & 103--116   & 102 & 413 & 22979.863 & 29707.296 & 6727.433 & 156289.48 \\
4aoi  & VVLPYMKHGDLRNF  & 14 & 1155--1168 & 102 & 413 & 23245.373 & 32378.950 & 9133.577 & 13328.65 \\
4cig  & VRDQAEHLKTAVQM  & 14 & 165--178   & 102 & 413 & 21375.594 & 29846.536 & 8470.942 & 17293.54 \\
4clj  & ILMELMAGGDLKSF  & 14 & 1194--1207 & 102 & 413 & 23968.789 & 30839.148 & 6870.358 & 56855.98 \\
4fp1  & PVHTAVGTVGTAPL  & 14 & 21--34     & 102 & 413 & 22564.107 & 30593.710 & 8029.604 & 9301.82 \\
4jpx  & DYLEAYGKGGVKA   & 13 & 154--166   & 92  & 373 & 16962.095 & 22231.950 & 5269.856 & 90422.62 \\
4jpy  & DYLEAYGKGGVKAK  & 14 & 154--167   & 102 & 413 & 23332.068 & 30779.295 & 7447.227 & 12918.78 \\
4tmk  & IEGLEGAGKTTARN  & 14 & 8--21      & 102 & 413 & 22590.207 & 29135.420 & 6545.212 & 199292.66 \\
5cqu  & RKLGRGKYSEVFE   & 13 & 43--55     & 92  & 373 & 17865.392 & 22801.515 & 4936.123 & 7620.94 \\
5nkb  & MIITEYMENGALDK  & 14 & 689--702   & 102 & 413 & 22570.674 & 31770.986 & 9200.312 & 9311.28 \\
6udv  & SLSRVMIHVFSDGV  & 14 & 245--258   & 102 & 413 & 24186.062 & 33350.850 & 9164.788 & 188397.35 \\
\bottomrule
\end{tabular}
}
\label{tab_l}
\end{table*}

\begin{table*}[htbp]
\footnotesize
\centering
\caption{M group proteins}
\resizebox{\linewidth}{!}{
\begin{tabular}{lllr|rrrrrr}
\toprule
PDB ID & Residue Sequence & Seq. Length & Residues & Qubits & Depth & Lowest Energy & Highest Energy & Energy Range & Exec. Time (s) \\
\midrule
1e2l  & AQITMGMPY       & 9  & 124--132   & 54  & 221 & 1509.665 & 2837.818 & 1328.153 & 12951.69 \\
1gx8  & SAPLRVYVE       & 9  & 36--44     & 54  & 221 & 1626.015 & 3053.529 & 1427.514 & 14080.77 \\
1m7y  & TAGATSANE       & 9  & 117--125   & 54  & 221 & 1420.378 & 2714.983 & 1294.604 & 12918.04 \\
1zsf  & LLDTGADDTV      & 10 & 23--32     & 63  & 257 & 4283.258 & 6023.888 & 1740.630 & 5674.54 \\
2avo  & LIDTGADDTV      & 10 & 23--32     & 63  & 257 & 4711.417 & 6788.627 & 2077.210 & 5709.81 \\
2bfq  & AFPAVSAGIYGC    & 12 & 136--147   & 82  & 333 & 11784.906 & 16384.379 & 4599.473 & 10361.37 \\
2bok  & EDACQGDSGG      & 10 & 188--197   & 63  & 257 & 4365.802 & 6164.745 & 1798.942 & 6145.18 \\
2qbs  & HCSAGIGRSGT     & 11 & 214--224   & 72  & 293 & 6691.571 & 9356.871 & 2665.300 & 13899.11 \\
2vwo  & EDACQGDSGG      & 10 & 188--197   & 63  & 257 & 4175.516 & 6533.564 & 2358.048 & 5812.72 \\
2xxx  & GAVEDGATMTFF    & 12 & 683--694   & 82  & 333 & 14199.993 & 18862.515 & 4662.522 & 14962.26 \\
3b26  & ELISNSSDAL      & 10 & 47--56     & 63  & 257 & 3768.807 & 6015.566 & 2246.759 & 5546.94 \\
3d83  & YLVTHLMGAD      & 10 & 103--112   & 63  & 257 & 4235.343 & 6119.164 & 1883.822 & 19833.57 \\
3vf7  & LLDTGADDTV      & 10 & 23--32     & 63  & 257 & 3975.024 & 6162.421 & 2187.398 & 5348.25 \\
4f5y  & GLAWSYYIGYL     & 11 & 158--168   & 72  & 293 & 6408.497 & 8858.596 & 2450.099 & 6157.46 \\
4mc1  & LLDTGADDTV      & 10 & 23--32     & 63  & 257 & 4092.236 & 6199.231 & 2106.996 & 5609.02 \\
4y79  & DACQGDSGG       & 9  & 189--197   & 54  & 221 & 1549.162 & 2874.211 & 1325.049 & 207445.70 \\
5cxa  & FDGKGGILAHA     & 11 & 174--184   & 72  & 293 & 6946.425 & 9298.822 & 2352.396 & 5638.71 \\
5kqx  & LLNTGADDTV      & 10 & 23--32     & 63  & 257 & 4336.777 & 6158.301 & 1821.524 & 21706.78 \\
5kr2  & LLNTGADDTV      & 10 & 23--32     & 63  & 257 & 4113.621 & 6383.194 & 2269.573 & 5687.63 \\
5nkc  & MIITEYMENGAL    & 12 & 689--700   & 82  & 333 & 12919.795 & 16929.422 & 4009.627 & 6363.43 \\
5nkd  & MIITEYMENGA     & 11 & 689--699   & 72  & 293 & 7192.774 & 10425.425 & 3232.651 & 5997.07 \\
6ezq  & AKQRLKCASL      & 10 & 194--203   & 63  & 257 & 4178.824 & 6002.270 & 1823.446 & 23591.38 \\
6g98  & RNNGHSVQLTL     & 11 & 60--70     & 72  & 293 & 7254.135 & 9951.906 & 2697.771 & 7080.74 \\
\bottomrule
\end{tabular}
}
\label{tab_m}
\end{table*}

\begin{table*}[htbp]
\footnotesize
\centering
\caption{S group proteins}
\resizebox{\linewidth}{!}{
\begin{tabular}{lllr|rrrrrr}
\toprule
PDB ID & Residue Sequence & Seq. Length & Residues & Qubits & Depth & Lowest Energy & Highest Energy & Energy Range & Exec. Time (s) \\
\midrule
1e2k  & DGPHGM       & 6  & 55--60    & 23 & 97  & 97.347   & 392.073  & 294.726 & 4425.19 \\
1hdq  & SIHSYS       & 6  & 194--199  & 23 & 97  & 135.525  & 400.060  & 264.535 & 4352.49 \\
1ppi  & PWWERYQP     & 8  & 57--64    & 46 & 189 & 1843.649 & 2795.853 & 952.204 & 13305.89 \\
1qin  & QQTMLRV      & 7  & 32--38    & 38 & 157 & 258.484  & 775.731  & 517.247 & 19567.41 \\
2v25  & ATFTIT       & 6  & 81--86    & 23 & 97  & 100.416  & 340.832  & 240.416 & 22356.46 \\
3ckz  & VKDRS        & 5  & 149--153  & 12 & 53  & 10.433   & 14.651   & 4.218   & 5763.36 \\
3dx3  & HNDPGWI      & 7  & 90--96    & 38 & 157 & 339.992  & 962.620  & 622.628 & 4661.24 \\
3eax  & RYRDV        & 5  & 45--49    & 12 & 53  & 10.357   & 16.021   & 5.664   & 4028.72 \\
3ibi  & IQFHFH       & 6  & 91--96    & 23 & 97  & 120.664  & 455.422  & 334.758 & 4486.62 \\
3nxq  & VCHASAWD     & 8  & 329--336  & 46 & 189 & 1815.928 & 2836.486 & 1020.558 & 14496.99 \\
3s0b  & GIKAVM       & 6  & 67--72    & 23 & 97  & 162.239  & 431.986  & 269.747 & 51428.83 \\
3tcg  & IEGVPESN     & 8  & 57--64    & 46 & 189 & 1660.359 & 2492.704 & 832.345 & 4331.88 \\
4mo4  & NIGGF        & 5  & 162--166  & 12 & 53  & 10.636   & 16.117   & 5.480   & 25834.89 \\
4q87  & SLTTPPLL     & 8  & 197--204  & 46 & 189 & 1659.516 & 2928.576 & 1269.061 & 4565.00 \\
4xaq  & GSYSDVSI     & 8  & 142--149  & 46 & 189 & 1486.347 & 2716.796 & 1230.450 & 4497.95 \\
4zb8  & GGPNGWKV     & 8  & 14--21    & 46 & 189 & 1791.084 & 2876.999 & 968.063 & 16029.02 \\
5c28  & CDLCSVT      & 7  & 663--669  & 38 & 157 & 386.810  & 792.776  & 405.965 & 114029.96 \\
5tya  & SLTTPPLL     & 8  & 197--204  & 46 & 189 & 1719.112 & 2594.339 & 875.227 & 9870.15 \\
6czf  & LRKANG       & 6  & 44--49    & 23 & 97  & 114.701  & 376.059  & 261.358 & 4309.82 \\
6p86  & VYSSGIPL     & 8  & 300--307  & 46 & 189 & 1486.200 & 3008.481 & 1522.281 & 4290.98 \\
\bottomrule
\end{tabular}
}
\label{tab_s}
\end{table*}

\textbf{Docking Results:} Docking simulations for each predicted structure were performed using AutoDock Vina~\cite{trott2010autodock}. To ensure result robustness, each structure underwent 20 independent docking simulations, each initialized with a distinct random seed. Each simulation generated the top 10 binding poses along with their corresponding binding affinity scores. These results, along with their averaged binding scores, are stored in separate JSON files to facilitate direct comparison of prediction quality among different modeling methods.

This structured approach ensures that \textit{QDockBank} is user-friendly, reproducible, and easily accessible to researchers from both quantum computing and computational biology communities.

\subsection{Protein Prediction of \textit{QDockBank}}

\begin{figure}[htbp]
    \centering
    \includegraphics[width=0.85\linewidth]{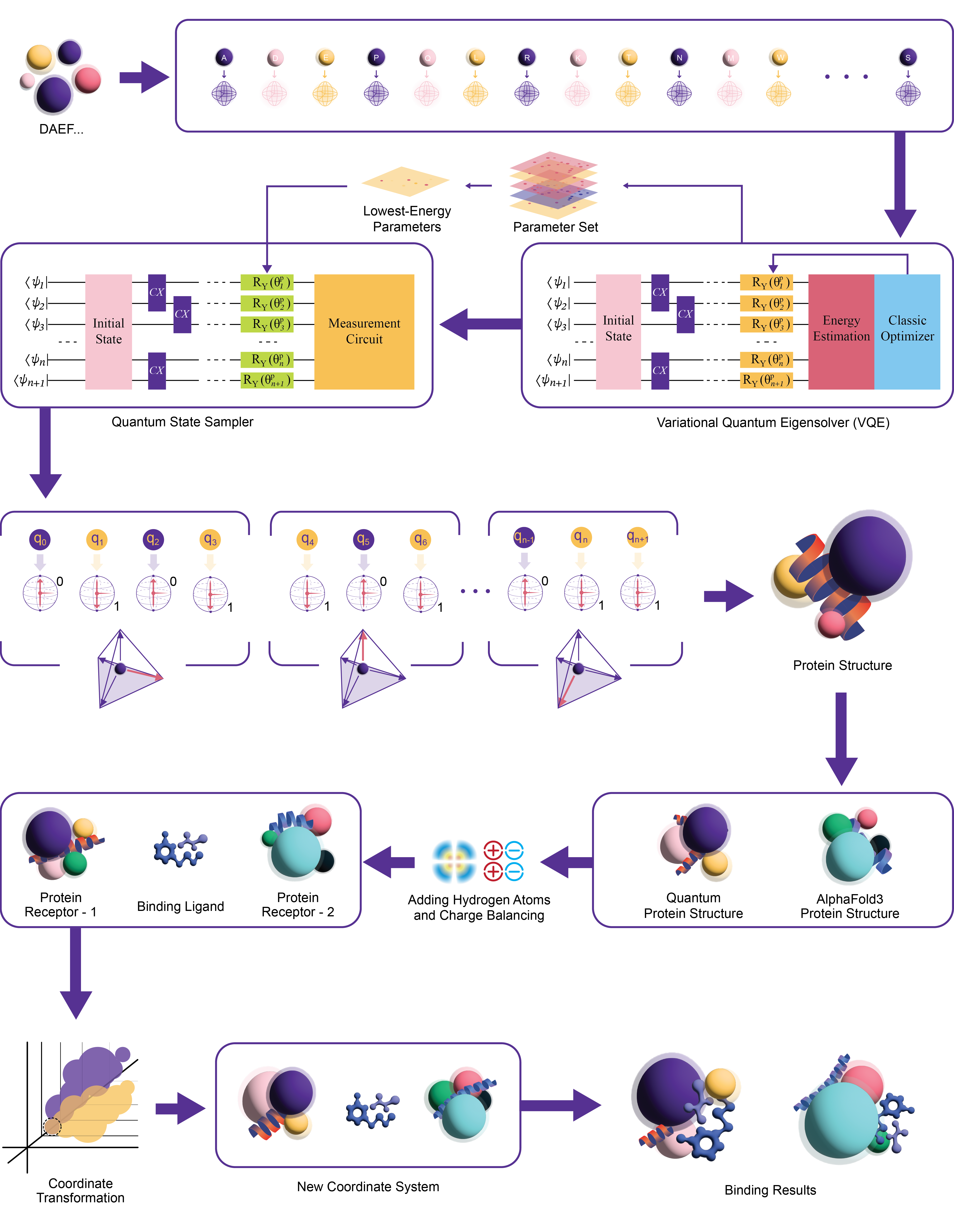}
    \caption{Overview of the protein fragment structure prediction workflow based on VQE, including lattice encoding, Hamiltonian construction, quantum optimization, and docking evaluation.}
    \label{fig2}
\end{figure}

Our prediction workflow is illustrated in Figure~\ref{fig2}. We employ a staged quantum engineering architecture based on a quantum-classical hybrid approach to predict structures of protein fragments, with a particular focus on ligand-binding active sites.

\subsubsection{Problem Mapping and Quantum Encoding}

Given a protein sequence fragment as input, we construct a coarse-grained representation using a tetrahedral lattice model. Each residue is represented as a node with four possible extension directions, fixed bond length, and a bond angle of approximately $109.4^\circ$, capturing the stereochemical properties of the protein backbone~\cite{yue2024integration}. This encoding facilitates a complete mapping of the protein’s conformational energy landscape onto a quantum system. Each protein conformation corresponds to a quantum state, with its energy directly represented by the expectation value of a quantum Hamiltonian. Optimizing this quantum system to its ground state identifies the lowest-energy protein conformation. Physically, this variational search mirrors conformational sampling, where the lowest-energy state corresponds to the most stable structure.

Based on this encoding, we construct a Hamiltonian representing the energy landscape of the protein conformational space. The Hamiltonian is composed of four distinct constraint terms and is formulated as:
\[
H_t = \lambda_c H_c + \lambda_g H_g + \lambda_d H_d + \lambda_i H_i,
\]
where $H_c$ enforces chirality constraints, $H_g$ maintains geometric backbone constraints, $H_d$ prevents residue collisions, and $H_i$ encodes pairwise amino acid interaction energies. In our experiments, weighting coefficients $\lambda_c, \lambda_g, \lambda_d, \lambda_i$ are all set to 1; these can be adjusted to emphasize different structural properties based on specific research objectives.

\subsubsection{Variational Optimization Based on VQE}

We adopt the Variational Quantum Eigensolver (VQE)~\cite{tilly2022variational} framework to identify the ground state of the total Hamiltonian. During the optimization, a parameterized quantum circuit evolves the initial quantum state, while the classical optimization algorithm COBYLA minimizes the expectation value of the Hamiltonian:
\[
E(\boldsymbol{\theta}) = \langle \psi_i | U^\dagger(\boldsymbol{\theta}) H U(\boldsymbol{\theta}) | \psi_i \rangle.
\]
This expectation is iteratively estimated and minimized using COBYLA. Upon finding a near-optimal parameter set $\boldsymbol{\theta}^*$, we fix the quantum circuit and measure the final quantum state:
\[
| \psi(\boldsymbol{\theta}^*) \rangle = U(\boldsymbol{\theta}^*) | 0^{\otimes n} \rangle,
\]
obtaining measurement outcomes $x \in \{0,1\}^n$ with probabilities:
\[
p(x; \boldsymbol{\theta}^*) = |\langle x | \psi(\boldsymbol{\theta}^*)\rangle|^2.
\]

We employ the \texttt{EfficientSU2} ansatz from Qiskit, which effectively balances expressiveness and hardware compatibility for practical quantum computations~\cite{joshi2021evaluating}. The circuit comprises alternating layers of parameterized $R_yR_z$ rotations and entangling gates among adjacent qubits.

\subsubsection{Atomic Reconstruction and Docking Evaluation}

The predicted coarse-grained structures are refined by applying standard amino acid templates. Backbone atom positions are adjusted to standard bond lengths, and missing hydrogens and charges are incorporated using Open Babel~\cite{o2011open}. Structures are subsequently centered to facilitate docking procedures. Docking evaluations are performed using AutoDock Vina~\cite{trott2010autodock}, treating predicted structures from our method and other comparative approaches as rigid receptors. Multiple docking simulations with distinct random seeds are conducted to assess binding energy consistency and pose variability.

\section{Quantum Engineering Architecture}

\subsection{Quantum Platform}

Our experiments were conducted on IBM’s 127-qubit superconducting quantum processor~\cite{chow2021ibm}. Specifically, we utilized the IBM Eagle r3 processor, whose native gate set includes \texttt{ECR}, \texttt{ID}, \texttt{RZ}, \texttt{SX}, and \texttt{X} gates.

\subsection{Architecture for Batch Processing}

In prior studies, utility-level quantum processors have exhibited relatively low sensitivity to noise in protein structure prediction tasks~\cite{kim2023evidence}. Protein structure prediction involves navigating complex energy landscapes within high-dimensional conformational spaces, where classical algorithms frequently become trapped in local minima. Interestingly, moderate levels of quantum noise may function as stochastic perturbations, assisting in escaping local minima and enhancing global optimization performance. Despite this robustness, superconducting quantum hardware currently faces two principal limitations: short coherence times (e.g., $T_1 \approx 60$–$120\,\mu$s, $T_2 \approx 40$–$100\,\mu$s for IBM Eagle) and gate-level inaccuracies~\cite{siddiqi2021engineering, li2023error}. These limitations constrain achievable quantum circuit depth and total runtime, especially during repeated executions in batch-mode processing.

To address these challenges, we employed a coarse-grained modeling approach at the algorithmic level and performed over 200 iterations of gradient-free classical optimization on parameterized quantum circuits. Since the parameter updates are carried out classically, this method provides greater fault tolerance compared to fully quantum approaches such as the Quantum Approximate Optimization Algorithm (QAOA)~\cite{zhou2020quantum}. The quantum workflow executed on physical processors consists of two stages. In the first stage, the ground-state energy of the system is approximated without requiring high-precision measurements, thereby reducing coherence time demands. In the second stage, the optimized circuit parameters are fixed, and the circuit is repeatedly executed and measured 100,000 times. The resulting bitstrings, corresponding to low-energy conformations, are then mapped to their respective three-dimensional protein structures. This design enables robust batch execution of the entire \textit{QDockBank} dataset under conditions minimally affected by quantum noise.

\subsection{Quantum Circuit Margin Strategy}

For protein sequences requiring a relatively large number of qubits, we intentionally allocate additional qubits beyond those strictly required by the logical circuit. The primary objective of this strategy is to effectively reduce the overall depth of quantum circuits. In current quantum engineering practices, the common approach begins by generating an idealized quantum circuit representing the algorithm under ideal conditions. This circuit is subsequently compiled and mapped onto physical quantum processors using quantum computing frameworks such as Qiskit. However, physical qubits in real quantum hardware lack full connectivity. Consequently, certain logical qubit pairs, when mapped onto physical qubits, might lack direct connectivity~\cite{li2019tackling}. To overcome this issue, additional \texttt{SWAP} gates must be inserted to physically reposition qubits, enabling execution of the requisite entangling gates. As a result, the actual circuit depth on hardware can become significantly greater than that of the ideal circuit. In large-scale problems, the additional circuit depth induced by routing overhead can exacerbate decoherence effects, reducing quantum state fidelity. To mitigate this, we allocate an additional 5 to 10 ancilla qubits during execution. Recent studies indicate that a modest increase in the number of available qubits significantly reduces overall circuit depth by enhancing routing flexibility and minimizing the number of required SWAP operations~\cite{li2020qubit}.

\section{Dataset Evaluation}

\subsection{Evaluation Metrics}

To comprehensively evaluate the effectiveness and practicality of quantum-computing-based protein structure prediction in ligand docking tasks, we define the following evaluation metrics:

\subsubsection{Structural Prediction Accuracy (RMSD)}

We use Root Mean Square Deviation (RMSD) between the predicted structure and the experimentally determined X-ray crystallographic structure to assess prediction accuracy~\cite{wang2004pdbbind}. This metric quantifies the spatial deviation between predicted protein pocket fragments and their experimentally validated counterparts. Lower RMSD values indicate predictions closer to actual structures. For each predicted fragment, we compute RMSD relative to the corresponding experimental structure using Biopython~\cite{cock2009biopython, chapman2000biopython}. QDockBank compares each predicted structure with its experimentally determined counterpart from the PDBbind dataset~\cite{wang2004pdbbind}. Specifically, we calculate RMSD based on the coordinates of backbone $\alpha$-carbon atoms (C$\alpha$). The positions of these atoms represent the core scaffold of the protein, providing a meaningful basis for evaluating structural differences between predicted and experimental conformations~\cite{kuroda2016pushing}.

\subsubsection{Binding Affinity Score from Docking}

We use AutoDock Vina~\cite{trott2010autodock} to perform molecular docking between predicted protein structures and their respective ligands, obtaining binding affinity scores. Binding affinity reflects interaction strength between the ligand and protein pocket, serving as an essential metric to evaluate whether predicted structures retain biological relevance. Lower (more negative) affinity scores indicate more favorable binding interactions~\cite{gilson2007calculation}. To enhance reliability, each predicted fragment undergoes multiple docking trials. Each docking trial is initiated with a distinct random seed. AutoDock Vina performs docking calculations and returns the top 10 binding poses ranked by their affinity scores. The poses are ordered from best to worst, and structural deviations among these poses are reported by Vina.

\subsection{Evaluation Results}

\begin{figure}[htbp]
    \centering
    \begin{subfigure}[t]{0.48\linewidth}
        \centering
        \includegraphics[width=\linewidth]{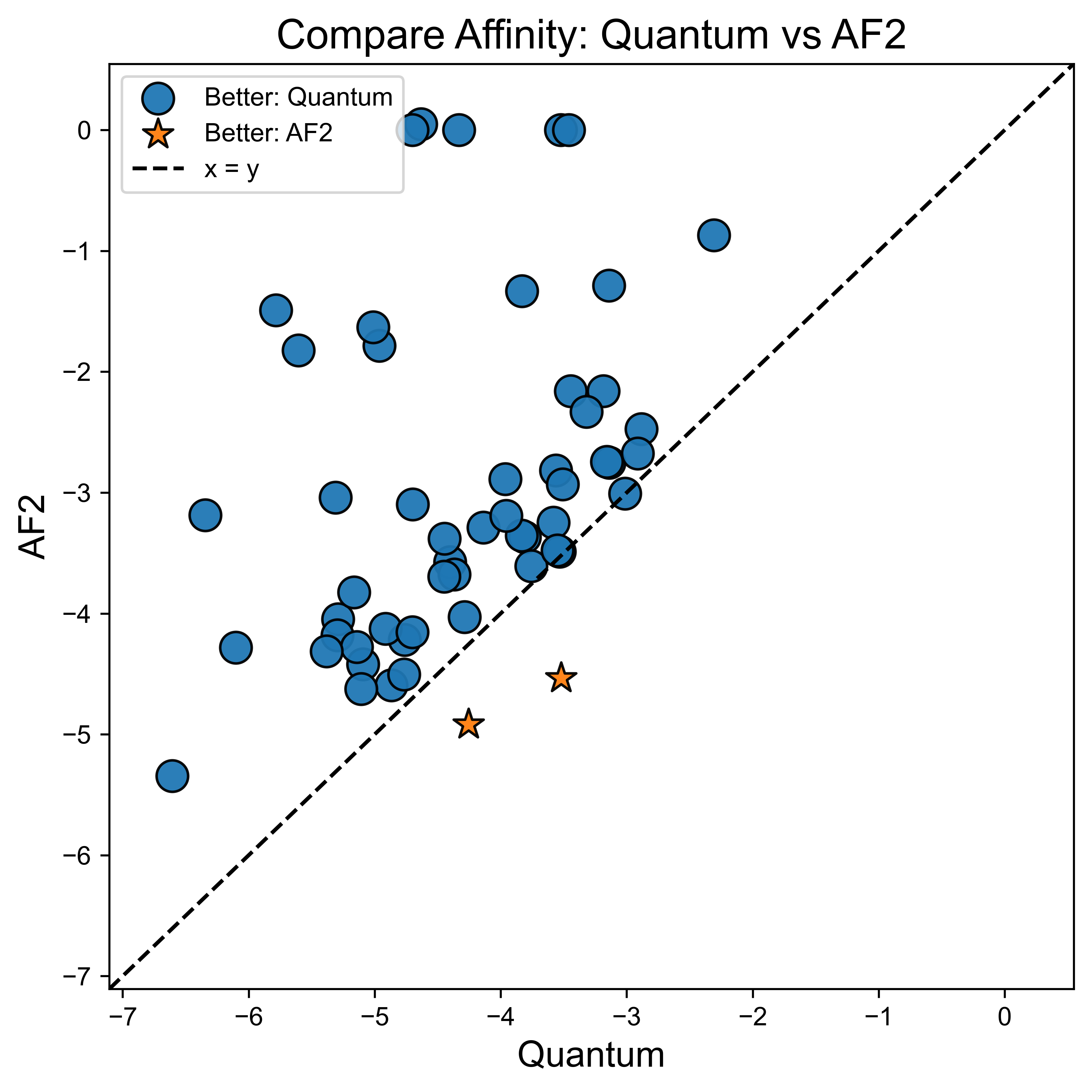}
        \caption{Affinity (All)}
        \label{fig:affinity_all}
    \end{subfigure}
    \hfill
    \begin{subfigure}[t]{0.48\linewidth}
        \centering
        \includegraphics[width=\linewidth]{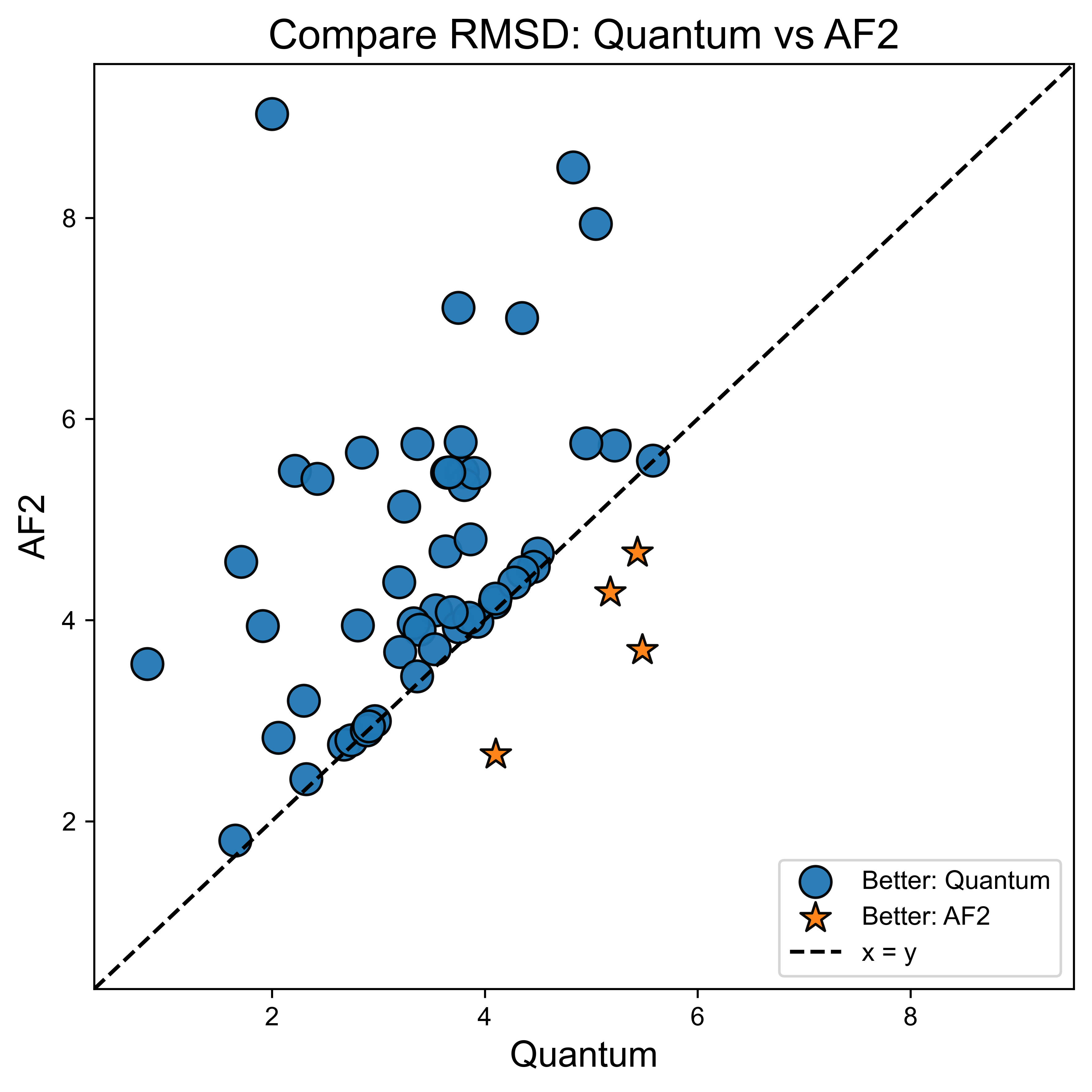}
        \caption{RMSD (All)}
        \label{fig:rmsd_all}
    \end{subfigure}
    \hfill
    \begin{subfigure}[t]{0.48\linewidth}
        \centering
        \includegraphics[width=\linewidth]{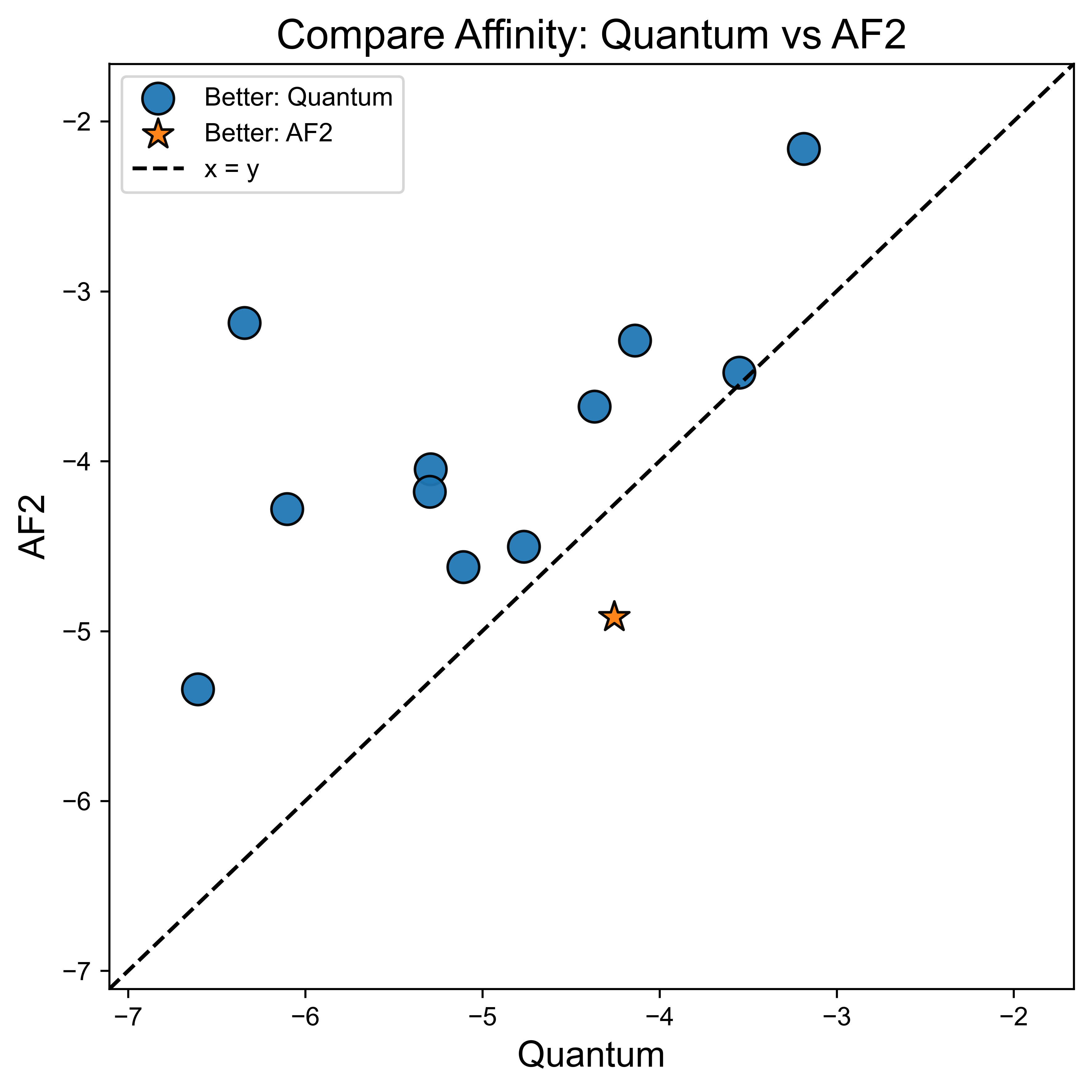}
        \caption{Affinity (L)}
        \label{fig:affinity_L}
    \end{subfigure}
    \hfill
    \begin{subfigure}[t]{0.48\linewidth}
        \centering
        \includegraphics[width=\linewidth]{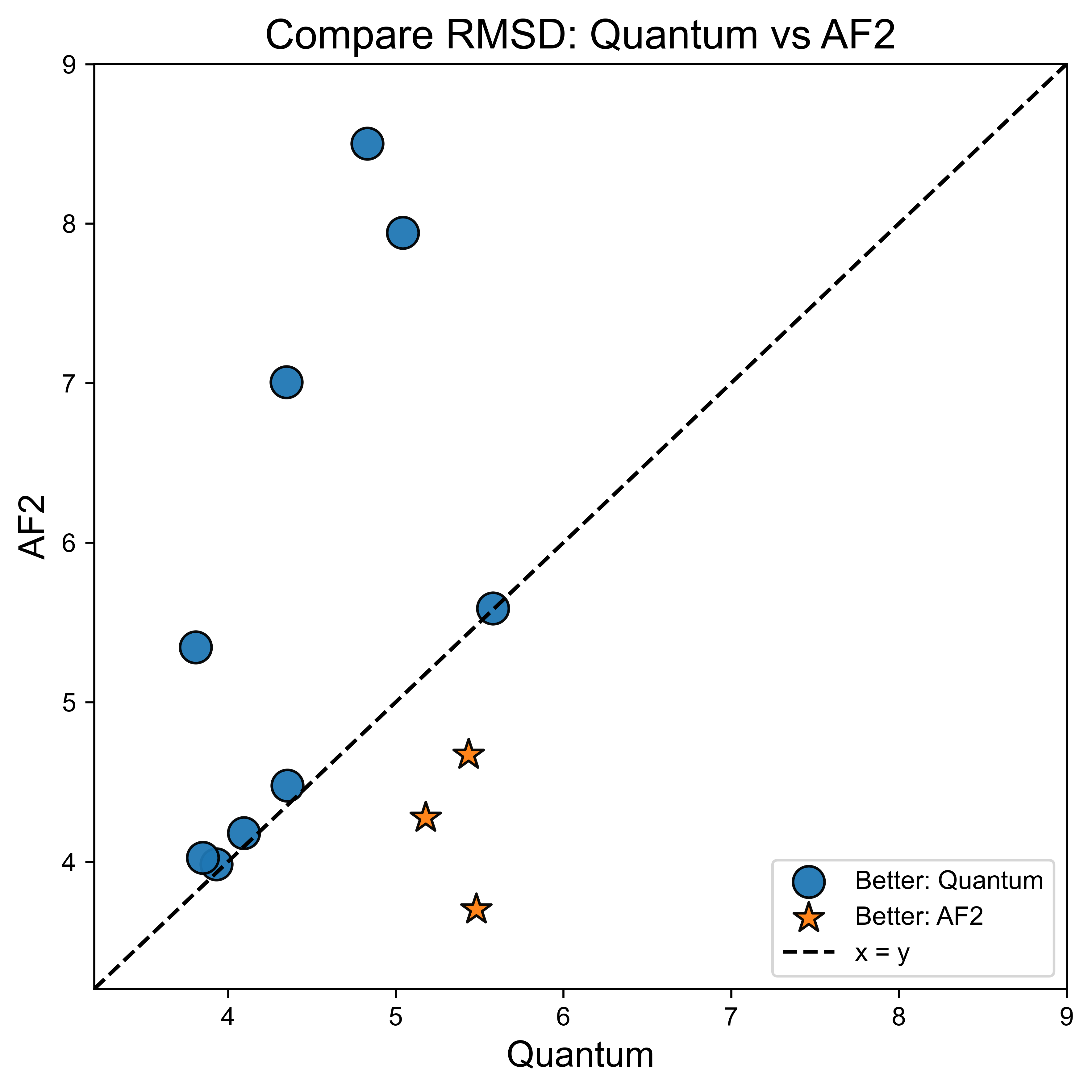}
        \caption{RMSD (L)}
        \label{fig:rmsd_L}
    \end{subfigure}
    \hfill
    \begin{subfigure}[t]{0.48\linewidth}
        \centering
        \includegraphics[width=\linewidth]{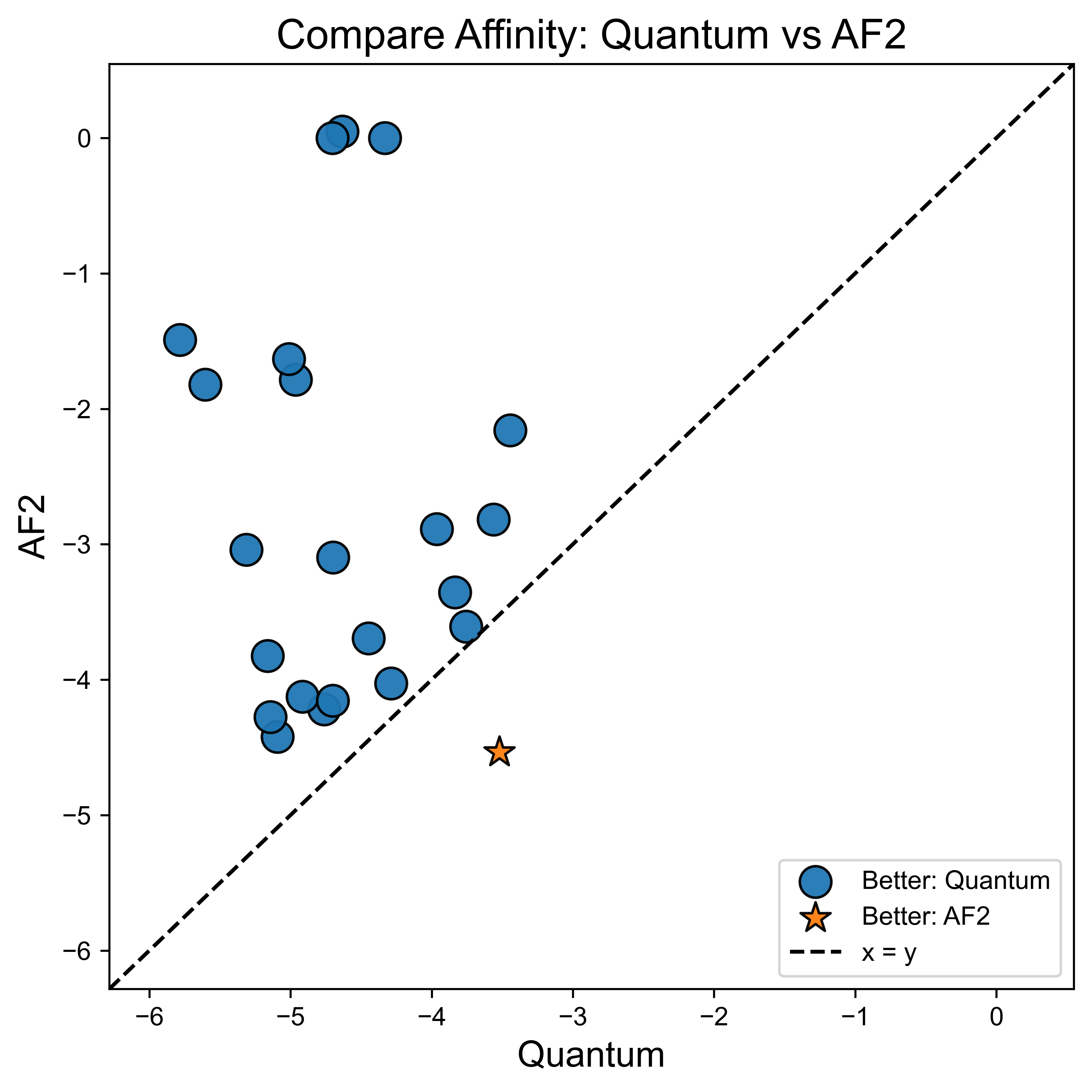}
        \caption{Affinity (M)}
        \label{fig:affinity_M}
    \end{subfigure}
    \hfill
    \begin{subfigure}[t]{0.48\linewidth}
        \centering
        \includegraphics[width=\linewidth]{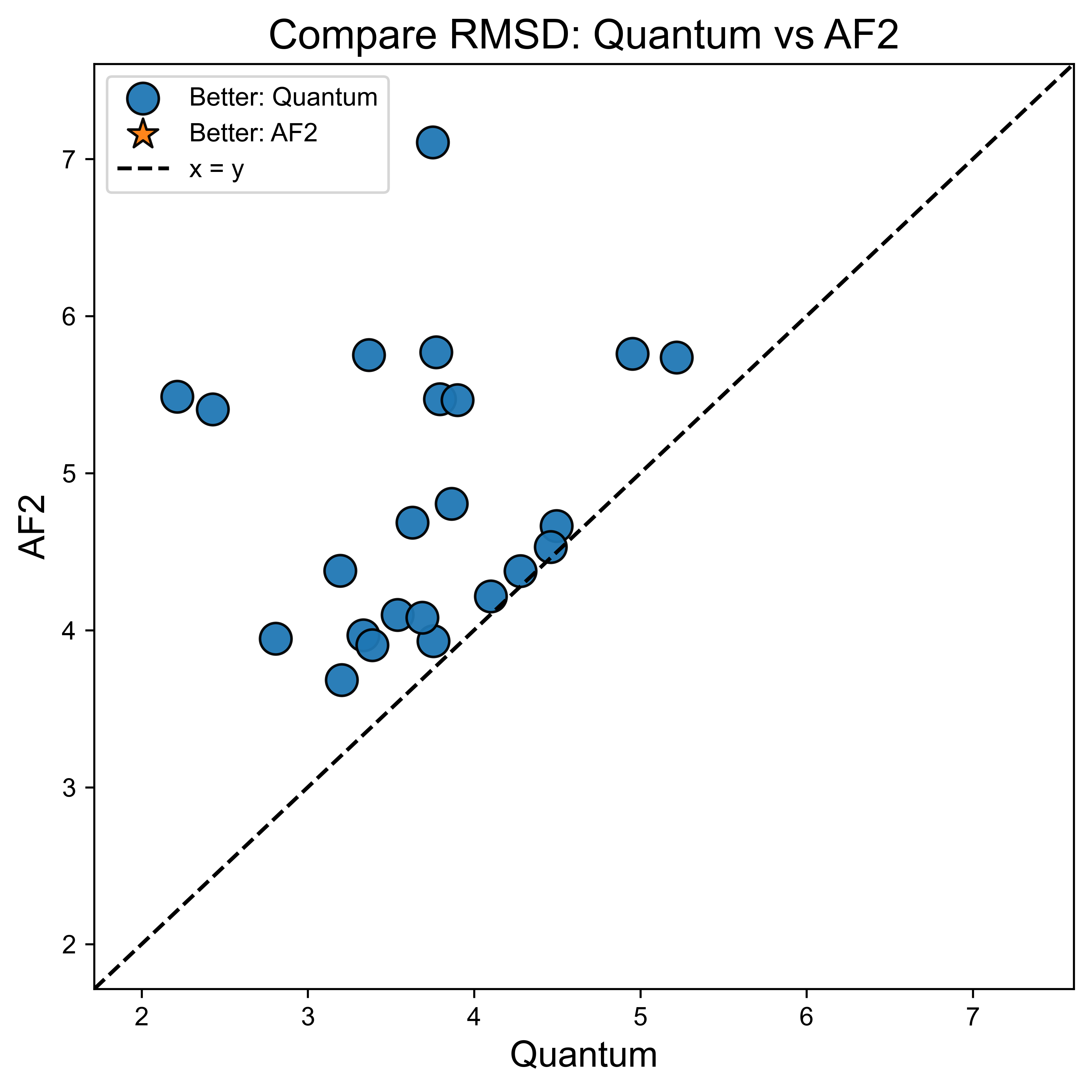}
        \caption{RMSD (M)}
        \label{fig:rmsd_M}
    \end{subfigure}
    \hfill
    \begin{subfigure}[t]{0.48\linewidth}
        \centering
        \includegraphics[width=\linewidth]{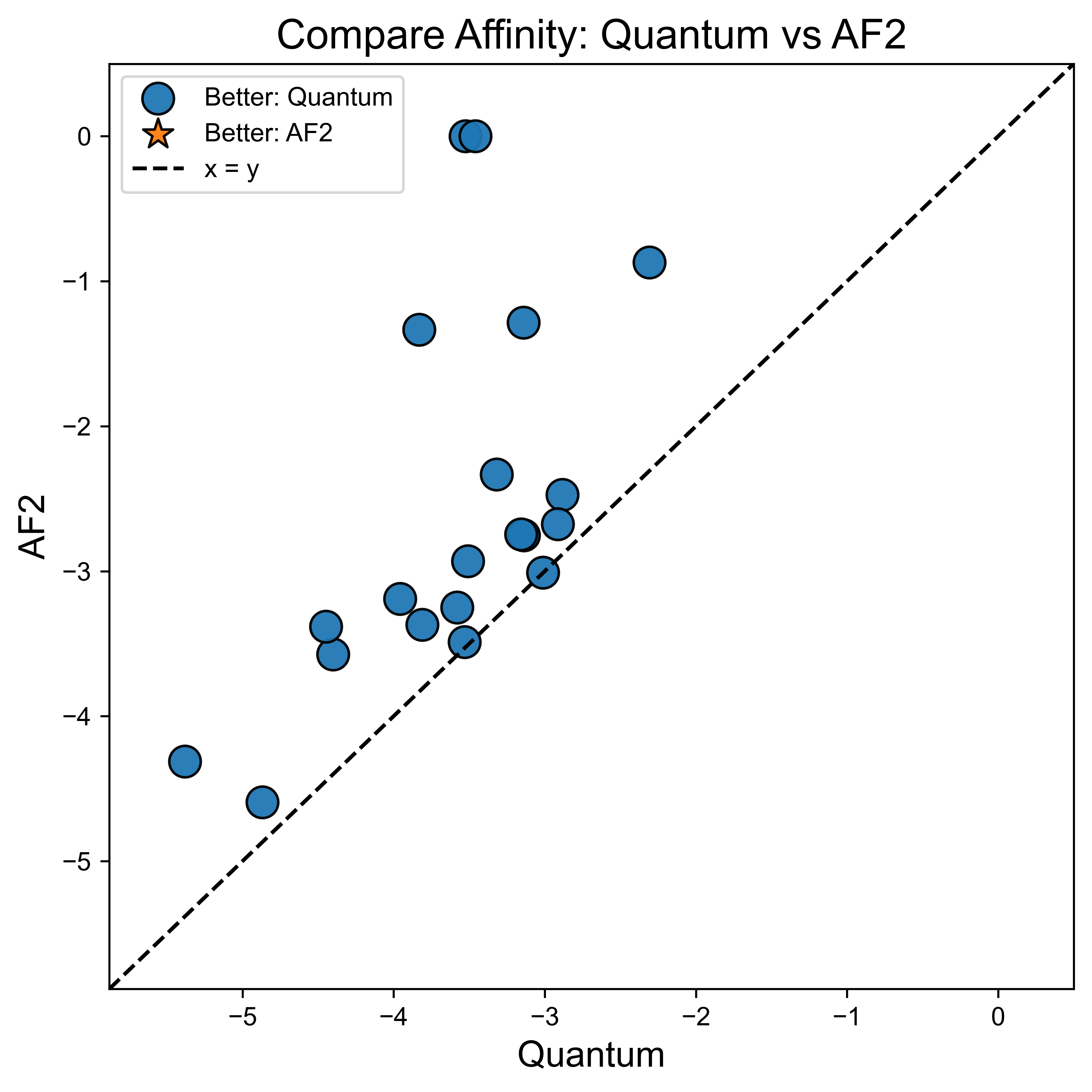}
        \caption{Affinity (S)}
        \label{fig:affinity_S}
    \end{subfigure}
    \hfill
    \begin{subfigure}[t]{0.48\linewidth}
        \centering
        \includegraphics[width=\linewidth]{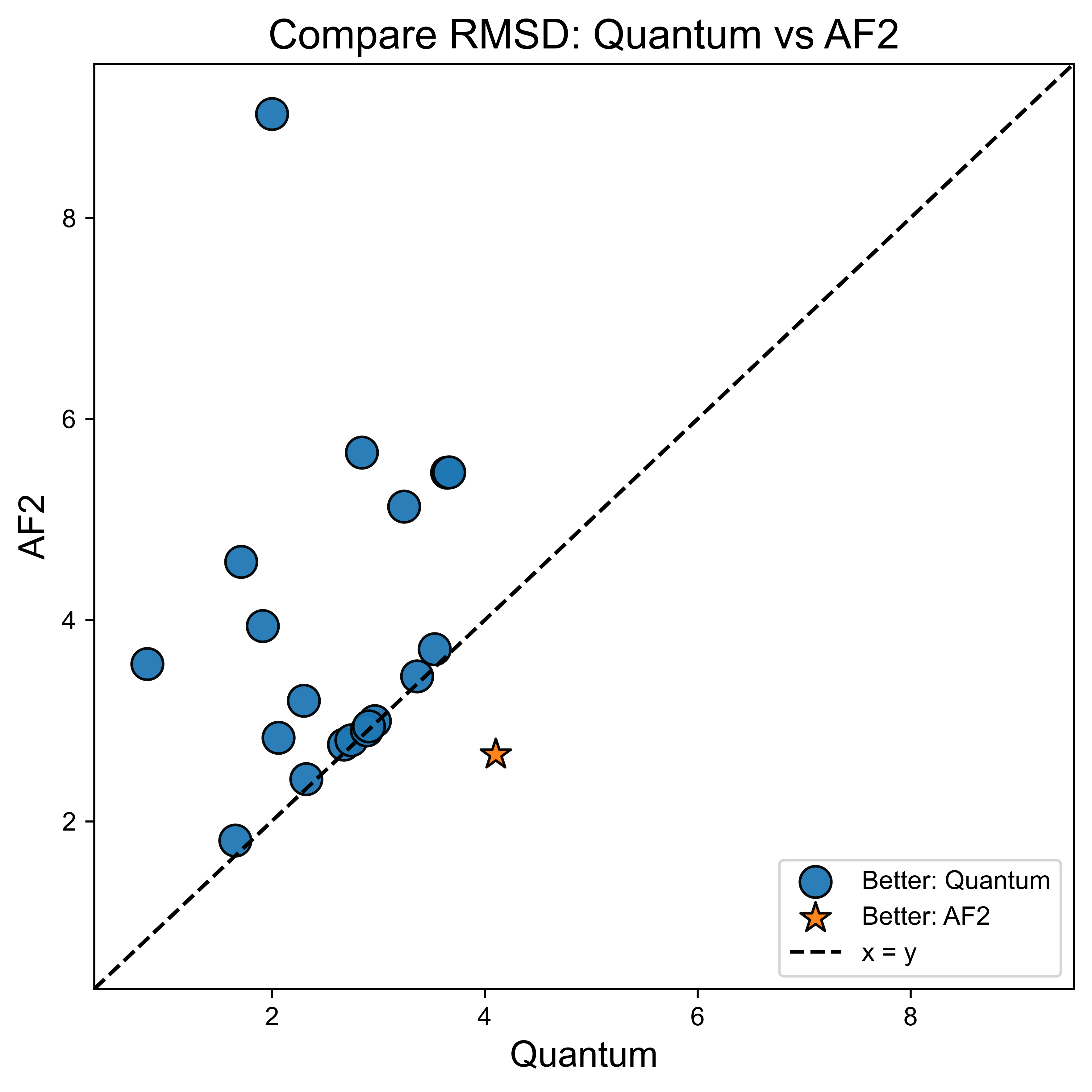}
        \caption{RMSD (S)}
        \label{fig:rmsd_S}
    \end{subfigure}

    \caption{Distribution of experimental results across all groups (All, L, M, S) between QDock and AF2. Each pair of plots illustrates the distribution of predicted docking binding affinity and RMSD for a specific group. Points above the diagonal indicate instances where QDockBank outperforms AF2, while points below the diagonal represent cases where AF2 performs better.}
    \label{fig3}
\end{figure}

\begin{figure}[htbp]
    \centering
    \begin{subfigure}[t]{0.48\linewidth}
        \centering
        \includegraphics[width=\linewidth]{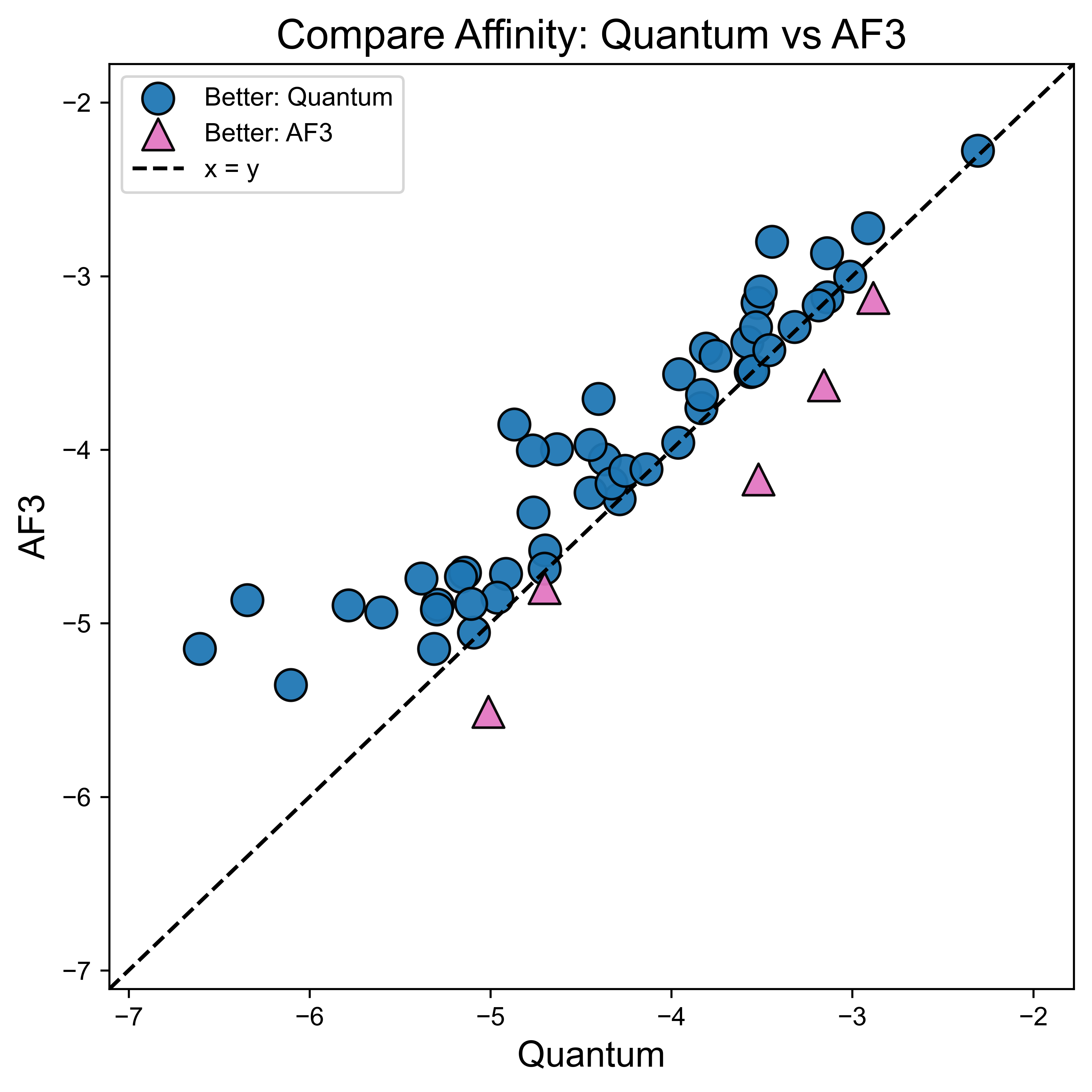}
        \caption{Affinity (All)}
        \label{fig:affinity_all}
    \end{subfigure}
    \hfill
    \begin{subfigure}[t]{0.48\linewidth}
        \centering
        \includegraphics[width=\linewidth]{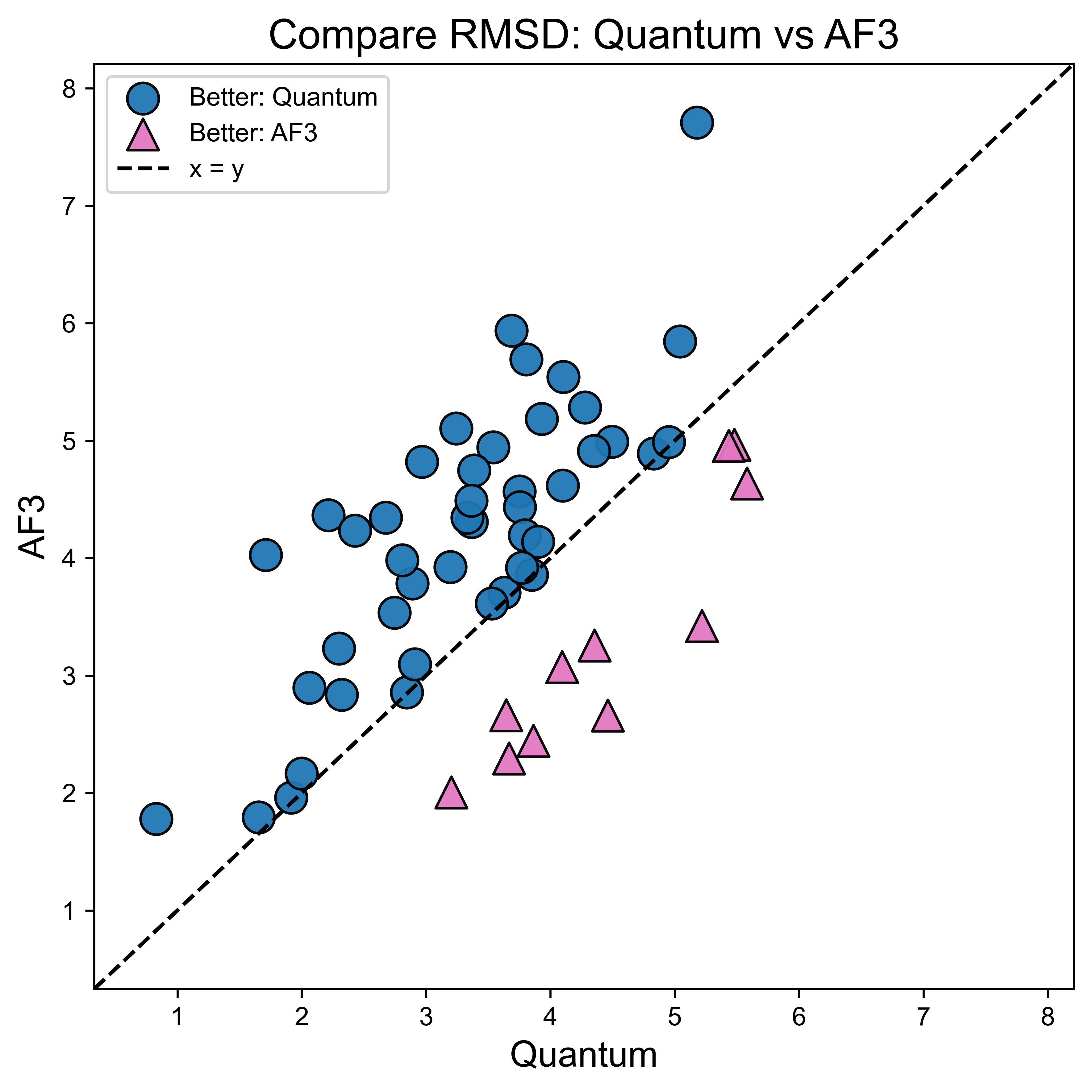}
        \caption{RMSD (All)}
        \label{fig:rmsd_all}
    \end{subfigure}
    \hfill
    \begin{subfigure}[t]{0.48\linewidth}
        \centering
        \includegraphics[width=\linewidth]{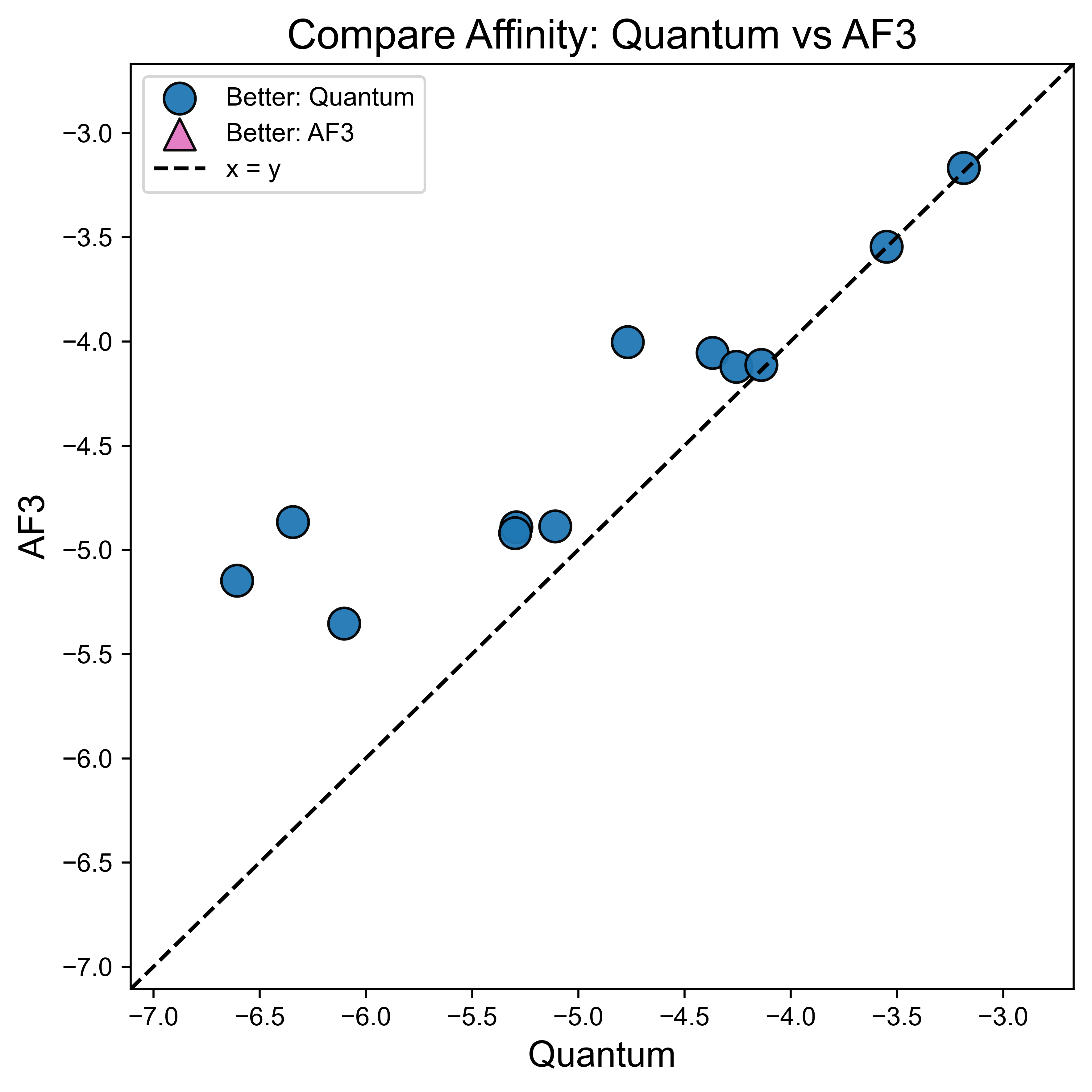}
        \caption{Affinity (L)}
        \label{fig:affinity_L}
    \end{subfigure}
    \hfill
    \begin{subfigure}[t]{0.48\linewidth}
        \centering
        \includegraphics[width=\linewidth]{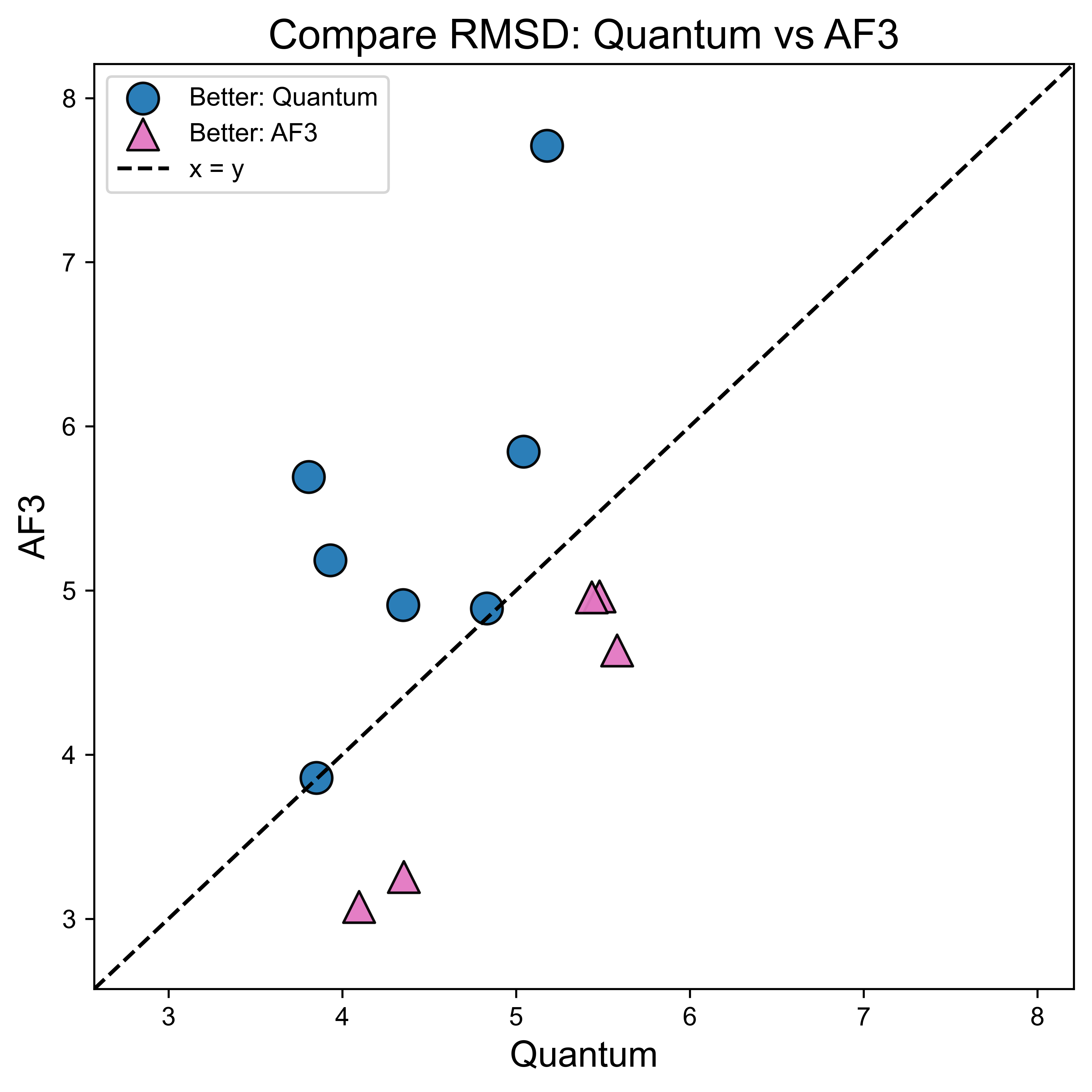}
        \caption{RMSD (L)}
        \label{fig:rmsd_L}
    \end{subfigure}
    \hfill
    \begin{subfigure}[t]{0.48\linewidth}
        \centering
        \includegraphics[width=\linewidth]{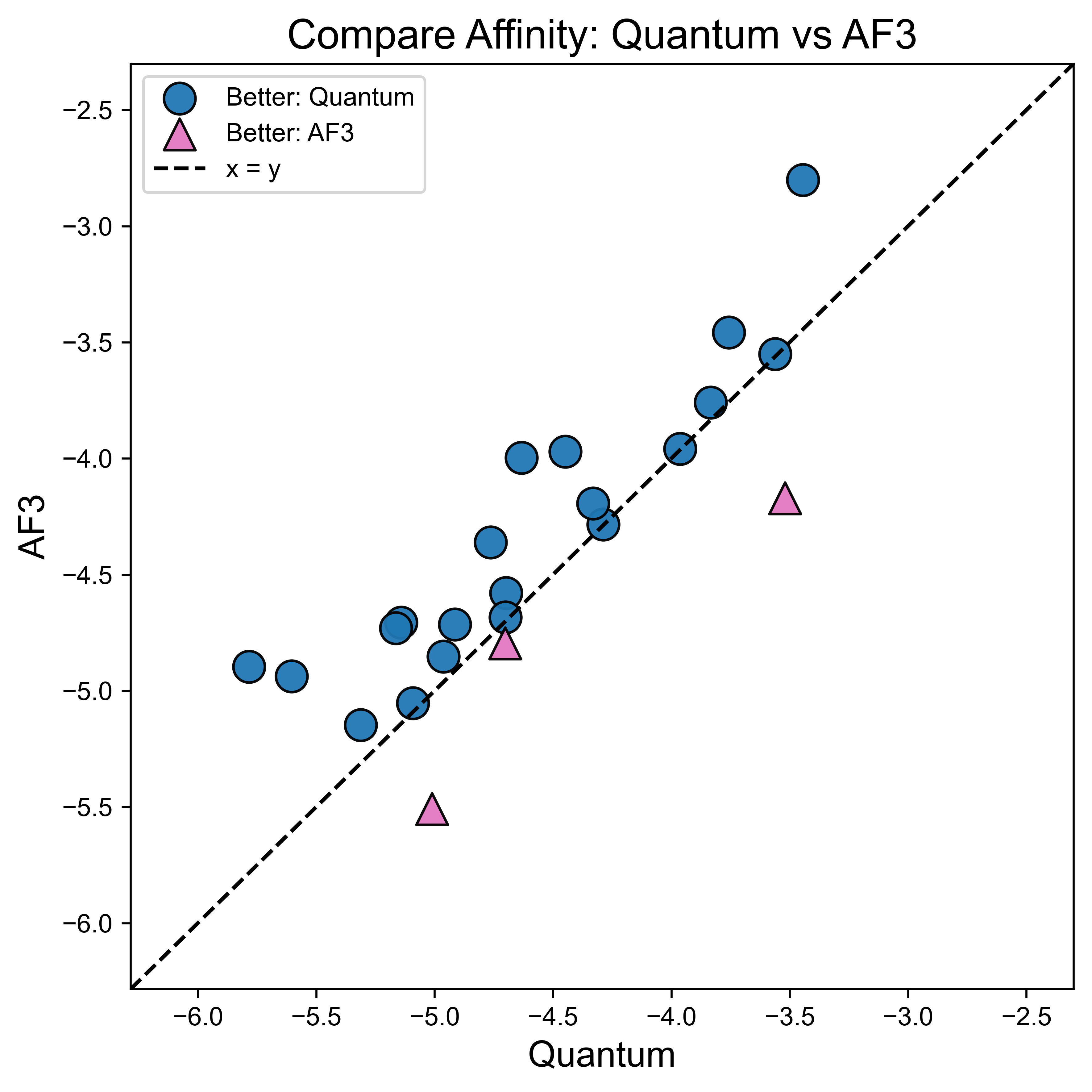}
        \caption{Affinity (M)}
        \label{fig:affinity_M}
    \end{subfigure}
    \hfill
    \begin{subfigure}[t]{0.48\linewidth}
        \centering
        \includegraphics[width=\linewidth]{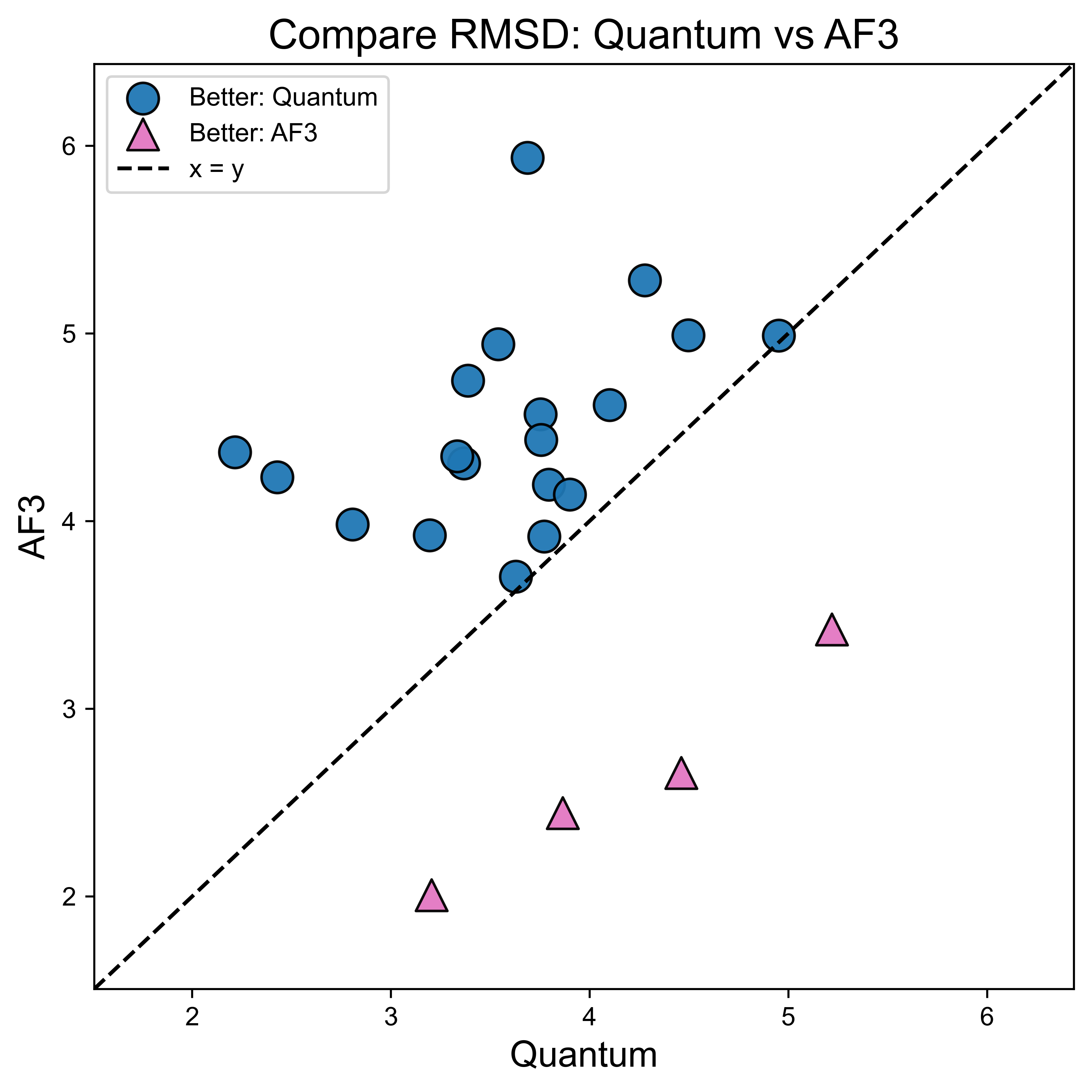}
        \caption{RMSD (M)}
        \label{fig:rmsd_M}
    \end{subfigure}
    \hfill
    \begin{subfigure}[t]{0.48\linewidth}
        \centering
        \includegraphics[width=\linewidth]{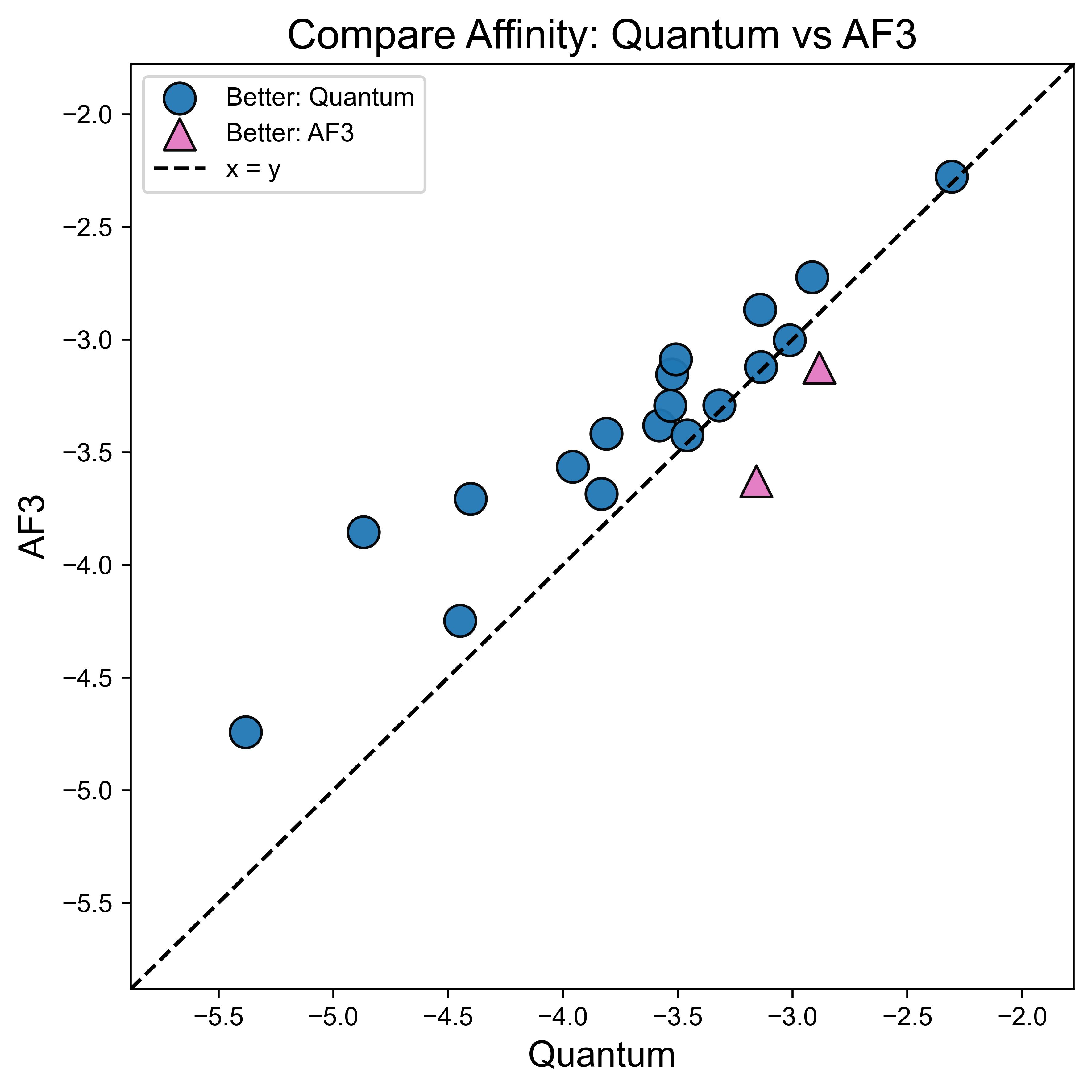}
        \caption{Affinity (S)}
        \label{fig:affinity_S}
    \end{subfigure}
    \hfill
    \begin{subfigure}[t]{0.48\linewidth}
        \centering
        \includegraphics[width=\linewidth]{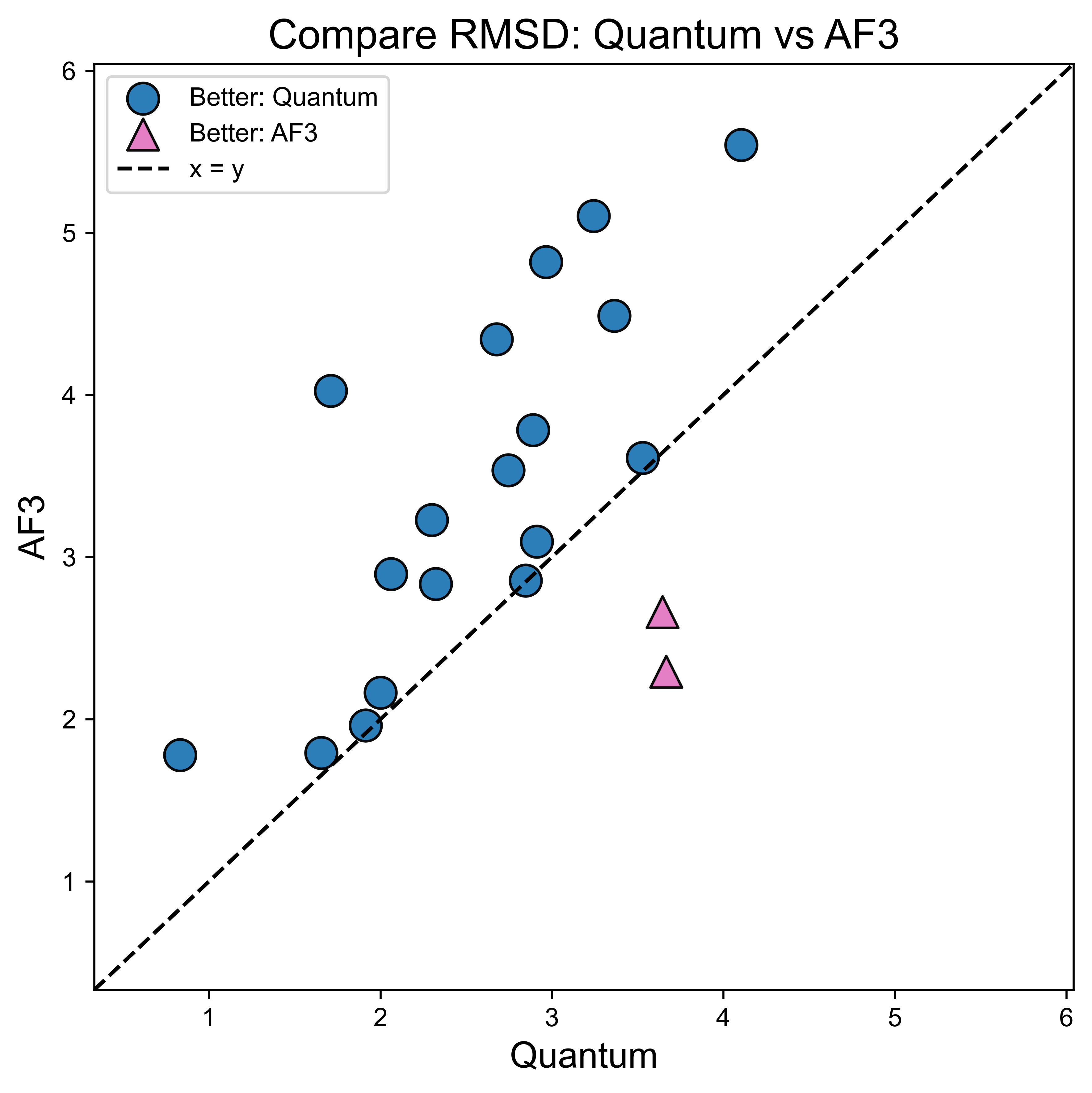}
        \caption{RMSD (S)}
        \label{fig:rmsd_S}
    \end{subfigure}

    \caption{Distribution of experimental results across all groups (All, L, M, S) between QDock and AF3. Each pair of plots illustrates the distribution of predicted docking binding affinity and RMSD for a specific group. Points above the diagonal indicate instances where QDockBank outperforms AF3, while points below the diagonal represent cases where AF3 performs better.}
    \label{fig4}
\end{figure}

\begin{figure*}[htbp]

    \centering
    \begin{subfigure}[t]{0.98\linewidth}
        \centering
        \includegraphics[width=\linewidth]{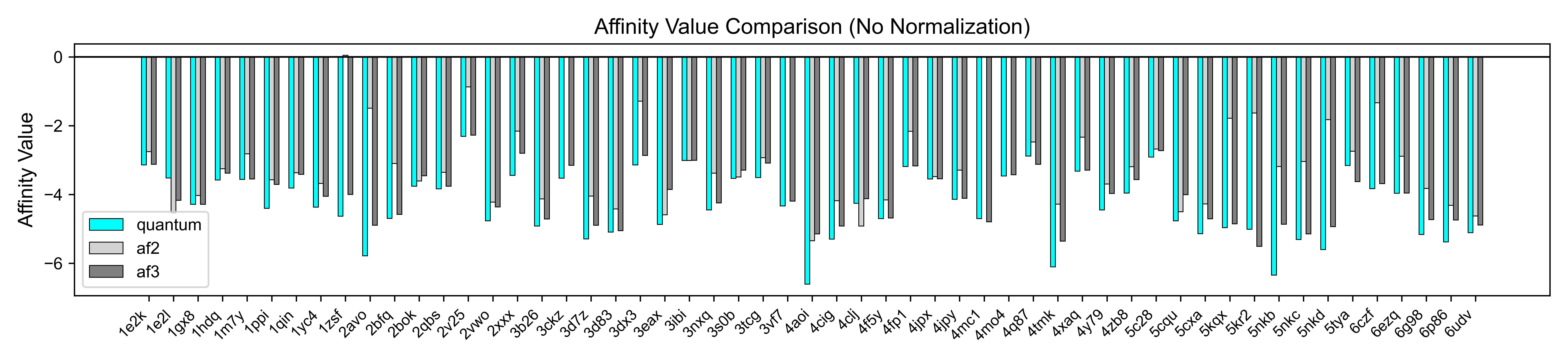}
        
    \end{subfigure}
    \hfill
    \begin{subfigure}[t]{0.98\linewidth}
        \centering
        \includegraphics[width=\linewidth]{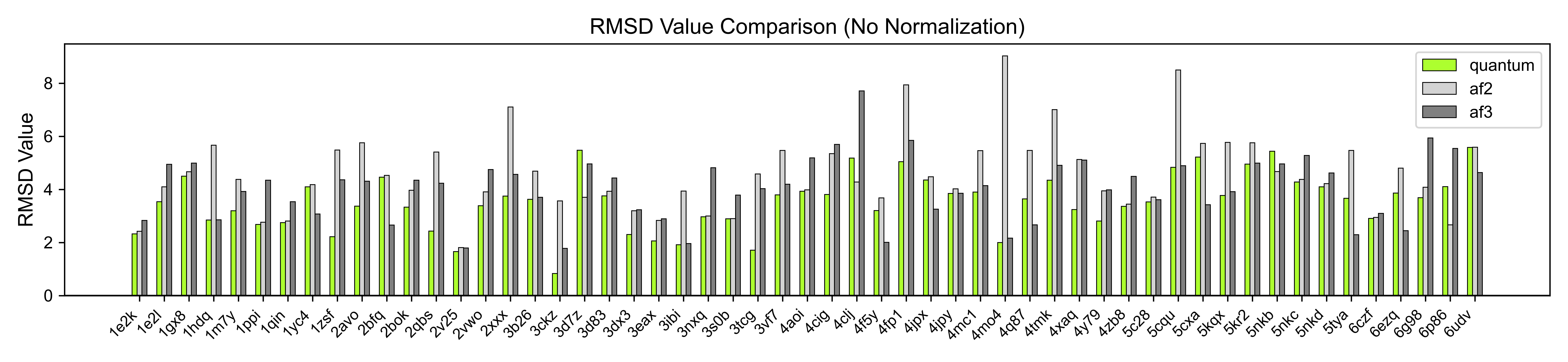}

    \end{subfigure}
    
    \caption{Evaluation metrics—affinity and RMSD—were analyzed within QDockBank and compared against results from AF2 and AF3. For both metrics, lower values indicate better performance. In these comparisons, the predicted structures from QDockBank outperformed those generated by the other two methods.}
    \label{fig5}
\end{figure*}

\textbf{Affinity and RMSD between predicted structures:} In each docking experiment, we dock the predicted protein fragment against its experimentally identified ligand from the PDBbind dataset and evaluate the resulting binding interactions. To ensure reproducibility, we record the random seed utilized in each docking simulation, thereby enabling reliable re-execution and validation of our docking outcomes. The distribution of experimental results is illustrated in Figures~\ref{fig3} and~\ref{fig4}. We present a side-by-side comparison across all methods and further group the samples by peptide length (S, M, L) to highlight the strengths of each method under different structural conditions. Figures~\ref{fig5} respectively present the statistical results of QDock compared with Alphafold2(AF2)~\cite{mirdita2022colabfold} and AF3.

Overall, our quantum-based approach shows better performance in terms of structural accuracy, particularly for medium (M) and long (L) peptide fragments. In several cases, the RMSD scores of our method are lower than those from AF2 and AF3. Regarding binding affinity, the quantum predictions also result in lower docking energy scores in many samples, indicating stronger ligand-binding plausibility. Our results, compared to those from AlphaFold2 (AF2), show that among all 55 data samples, QDock achieves better binding affinity scores in 53 cases, accounting for 96.4\% of the total. In Group L, QDock outperforms AF2 in 11 samples (91.7\%); in Group M, it shows better results in 22 cases (95.7\%); and in Group S, QDock achieves better scores in all 20 samples. Although QDock currently surpasses AF2 by a relatively small margin in individual docking cases, it demonstrates more stable performance and potential extensibility. In the RMSD comparison, QDock performs well in 51 of 55 samples, accounting for 92.7\%. Specifically, it performs better in 9 samples in Group L (75\%), in all 23 samples in Group M, and in 19 samples in Group S (95\%). Compared to AlphaFold3 (AF3), QDock achieves better binding affinity scores in 50 of 55 samples, corresponding to 90.9\%. In Group L, all 12 samples are better; in Group M, 20 samples show better scores (87\%); and in Group S, QDock performs better in 18 samples (90\%). However, in terms of RMSD comparison, AF3 begins to demonstrate superior performance. QDock still performs better in 40 samples in general (80\%), including 7 in Group L (58.7\%), 19 in Group M (82. 6\%) and 18 in Group S (90\%).

\begin{figure}[htbp]
    \centering
    \includegraphics[width=0.9\linewidth]{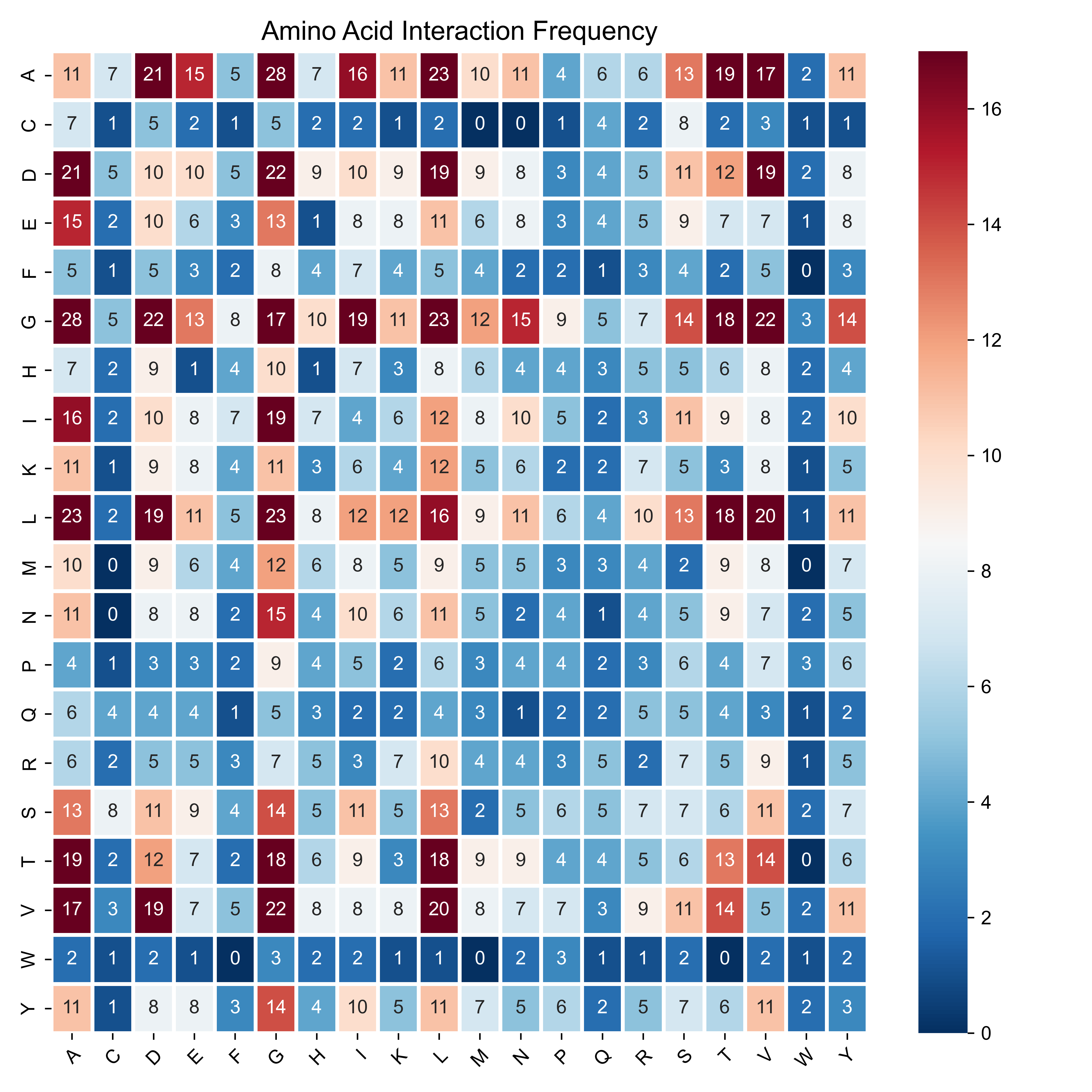}
    \caption{Frequency distribution of amino acid interactions in \textit{QDockBank}. QDockBank covers 395 out of the 400 possible amino acid interaction types, and common interaction pairs such as G–A and L–G appear with particularly high frequency in the dataset.}
    \label{fig6}
\end{figure}

\textbf{Amino acid interactions:} In addition to evaluating the structural accuracy of predicted conformations, another crucial aspect for practical applications is the coverage of amino acid interaction types within the dataset. According to protein free-energy minimization principles, the diversity and nature of amino acid–amino acid interactions significantly influence conformational preferences and the accuracy of energy estimations. These interactions include hydrophobic interactions, hydrogen bonds, electrostatic attractions and repulsions, and van der Waals forces. An incomplete representation of these interactions within a dataset can introduce biases and limit the generalizability of modeling outcomes. To ensure general applicability of \textit{QDockBank}, we systematically assessed the interaction coverage in the dataset. As illustrated in Figure~\ref{fig6}, \textit{QDockBank} covers approximately 98.75\% of all pairwise amino acid interactions observed in naturally occurring proteins. For further validation, we referred to the widely recognized Miyazawa–Jernigan interaction model, which defines a statistical energy matrix encompassing all 400 possible residue–residue combinations among the 20 standard amino acids~\cite{miyazawa1985estimation}. This model has been extensively utilized in coarse-grained modeling and energy-based structure predictions due to its biological relevance and statistical robustness. We confirmed that all residue–residue interactions defined by the Miyazawa–Jernigan model are present in \textit{QDockBank}, indicating full coverage of biologically relevant interaction types. This comprehensive representation ensures that the modeled energy landscapes accurately reflect genuine protein behavior, thus enhancing reliability in downstream applications such as energy function training, conformational sampling, and generalization assessments.

\textbf{Protein types:} Another critical factor when evaluating the performance of protein structure prediction models is the diversity of protein types included in the dataset. To ensure predictive models generalize beyond limited protein families, it is essential to incorporate a broad range of proteins varying in structure, function, and biological origin. Different protein classes often exhibit distinct structural motifs and physicochemical properties; while some, such as transmembrane helices and catalytic residues, are well characterized, others remain comparatively underexplored. Consequently, diversifying protein representation within the dataset facilitates rigorous performance benchmarking and promotes the discovery of novel structure–function relationships.

\textit{QDockBank} is primarily designed for structure-based docking tasks, focusing particularly on proteins with ligand-binding functions. The majority of fragments are derived from enzymes due to their central roles in catalysis, molecular recognition, and drug discovery. Specifically, the dataset encompasses the following diverse functional classes:
\begin{itemize}
    \item \textbf{Viral enzymes} (e.g., \texttt{1e2k}, \texttt{1e2l}, \texttt{1zsf}, \texttt{2avo}, \texttt{3vf7}), critical targets in antiviral drug development;
    \item \textbf{Kinases} (e.g., \texttt{3d7z}, \texttt{4aoi}, \texttt{4tmk}), regulatory proteins involved in signal transduction pathways;
    \item \textbf{Digestive and metabolic enzymes} (e.g., \texttt{1hdq}, \texttt{1m7y}, \texttt{3ibi}, \texttt{5cxa}), characterized by high structural diversity and metabolic importance;
    \item \textbf{Receptors and ligand-binding proteins} (e.g., \texttt{1gx8}, \texttt{3s0b}, \texttt{4xaq}), featuring complex binding surfaces critical for specificity;
    \item \textbf{Chaperones and regulatory proteins} (e.g., \texttt{1yc4}, \texttt{6udv}), involved in protein folding and cellular stress responses;
    \item \textbf{Proteases} (e.g., \texttt{5kqx}, \texttt{5kr2}), essential for protein degradation and cellular signaling processes;
    \item Additionally, a selection of \textbf{miscellaneous proteins} (e.g., \texttt{2bfq}, \texttt{5tya}) was included to further expand the dataset’s structural and functional heterogeneity.
\end{itemize}

Moreover, the dataset integrates proteins from diverse biological origins—including viruses, animals, and plants—to maximize heterogeneity. Such extensive coverage not only supports accurate structural predictions but also provides a robust, realistic environment for simulating structure-based docking scenarios relevant to drug discovery and bioengineering applications.

\section{Applications of QDockBank}

In this section, we highlight two principal applications of \textit{QDockBank}: protein–ligand docking and RMSD-based structural evaluation. These tasks demonstrate that \textit{QDockBank}'s quantum-predicted structures can be directly applied in downstream workflows with the minimal required preprocessing.

\subsection{Protein–Ligand Docking Readiness}

\begin{figure}[htbp]
    \centering
    \begin{subfigure}[t]{0.3\linewidth}
        \centering
        \includegraphics[trim=60 0 60 10, clip, width=\linewidth]{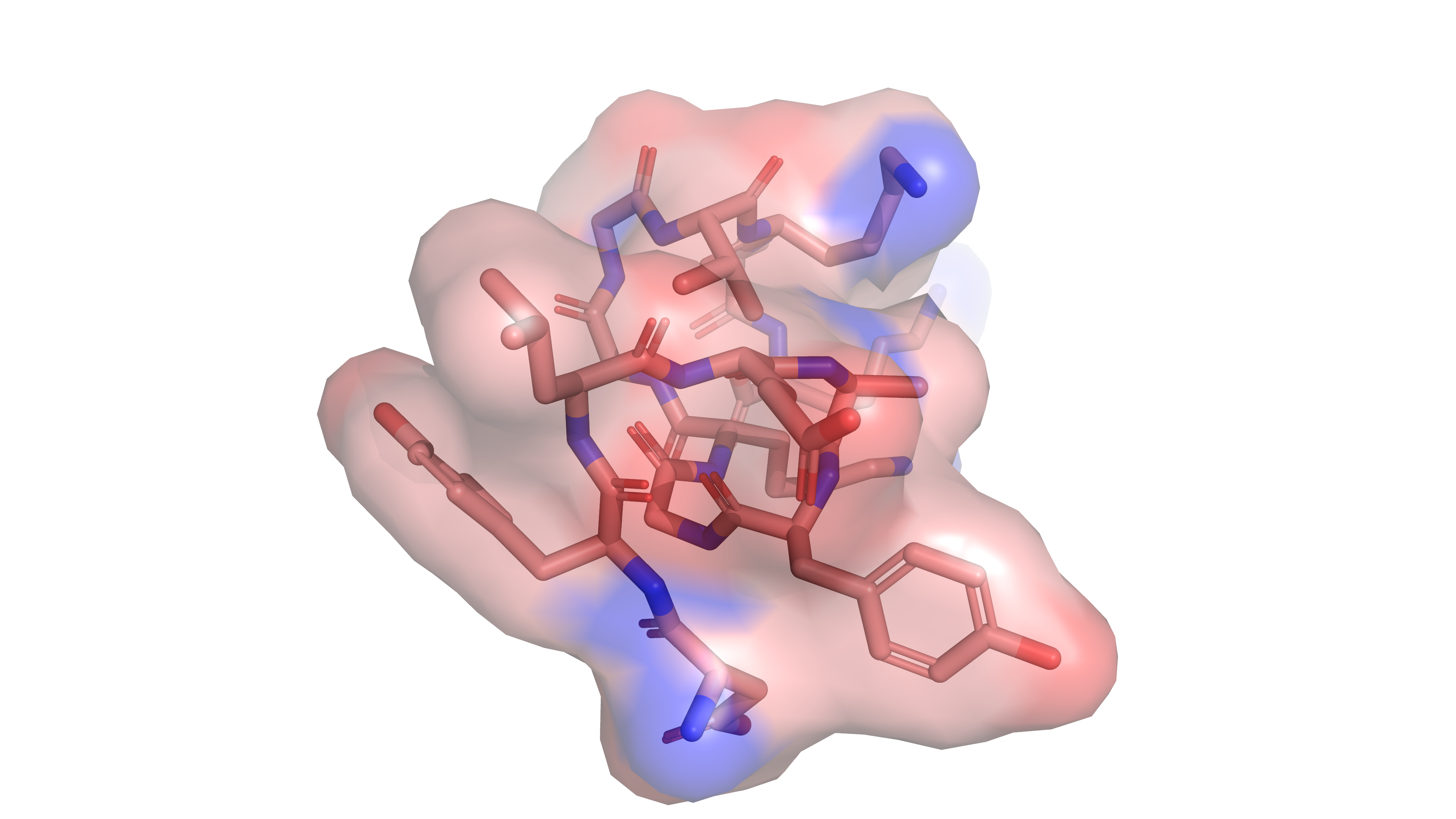}
        \caption{Protein}
    \end{subfigure}
    \hspace{0.01\linewidth}
    \begin{subfigure}[t]{0.3\linewidth}
        \centering
        \includegraphics[trim=60 10 60 20, clip, width=\linewidth]{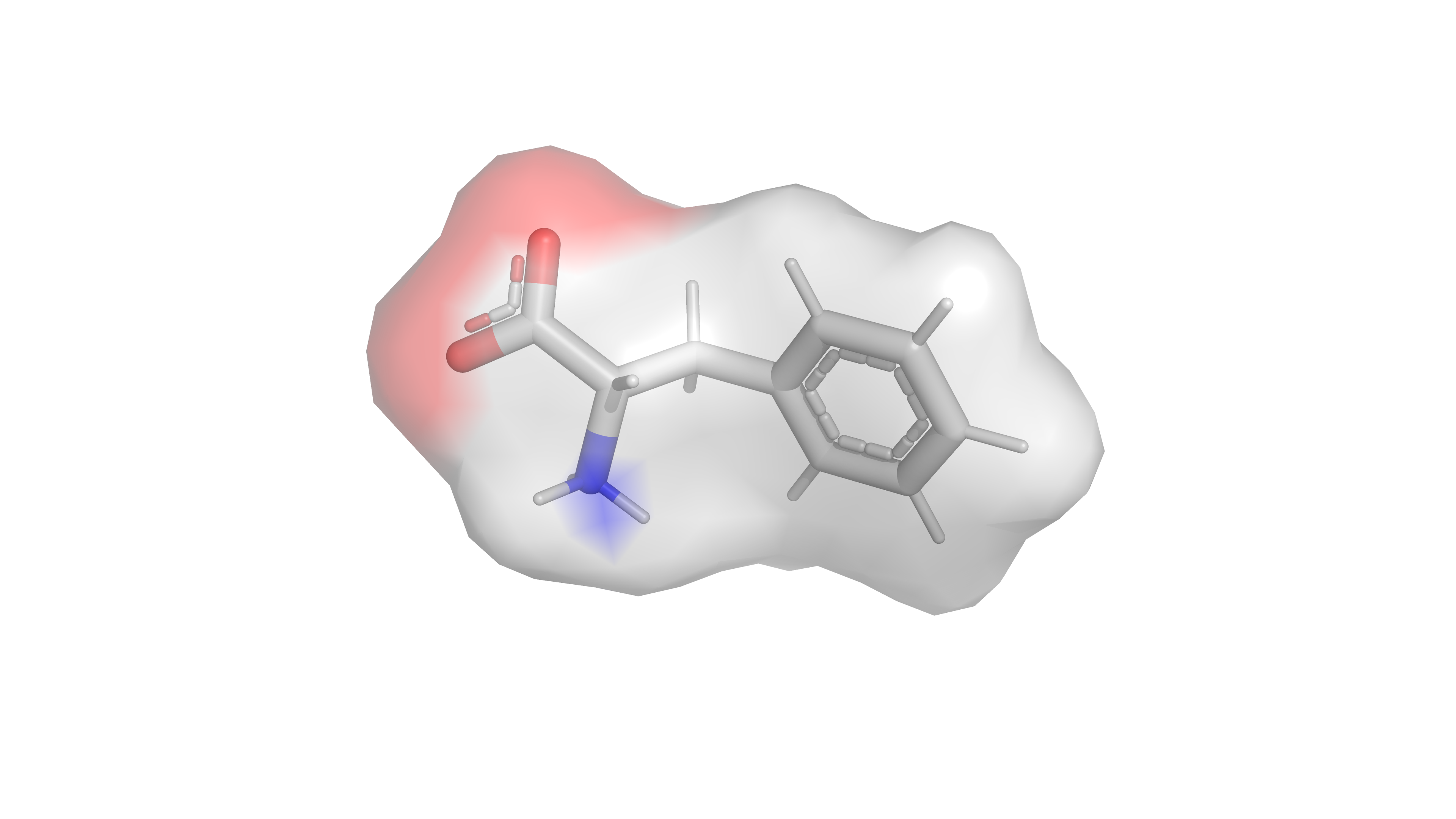}
        \caption{Ligand}
    \end{subfigure}
    \hspace{0.01\linewidth}
    \begin{subfigure}[t]{0.3\linewidth}
        \centering
        \includegraphics[trim=50 20 50 20, clip, width=\linewidth]{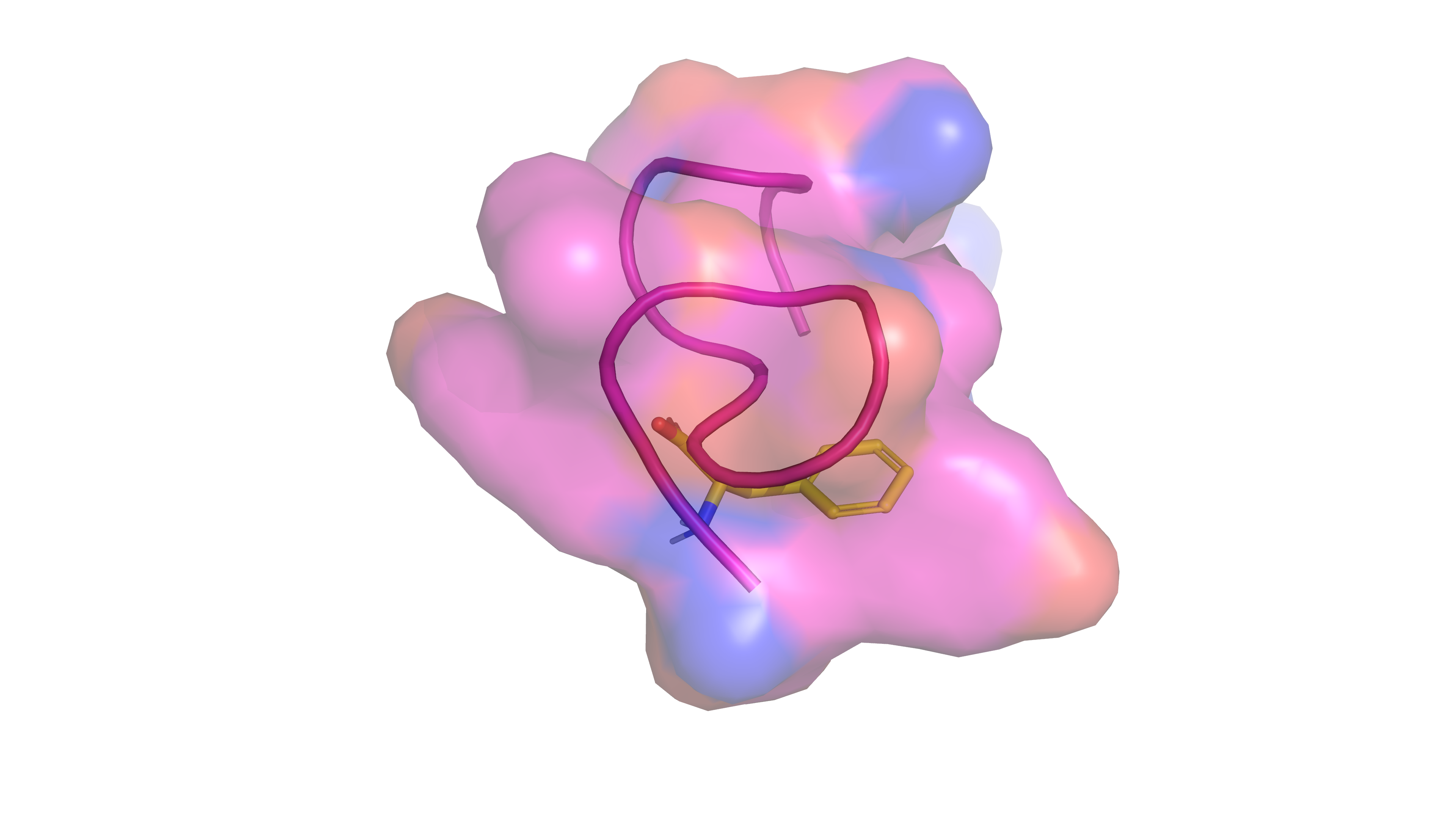}
        \caption{Docking result (combined)}
    \end{subfigure}
    
    \caption{Docking case for PDB entry \texttt{4jpy}: predicted fragment, ligand structure, and docking overlay. The visualized docking results show that the ligand is accurately embedded into the surface of the QDockBank-predicted protein structure, making direct contact with protein atoms in a manner that closely aligns with biologically relevant binding behavior.}
    \label{fig:docking_4jpy}
\end{figure}

All predicted protein fragments in \textit{QDockBank} are provided in standard PDB format with complete backbone atomic coordinates. These structures can be readily converted into the PDBQT format required by docking software such as AutoDock and AutoDock Vina using tools like AutoDockTools or Open Babel. In contrast to other methods that often require structural refinement prior to docking, \textit{QDockBank} facilitates direct integration into docking pipelines, substantially reducing the implementation barrier for practical use. To validate this applicability, we present docking case studies for PDB entries \texttt{4jpy}. For each case, the predicted fragment, corresponding native ligand, and final docking result are visualized (Figures~\ref{fig:docking_4jpy}). In case studies, the quantum-predicted structures exhibit chemical validity and spatial compatibility with their associated ligands.

As shown in Figure~\ref{fig:docking_4jpy}, the predicted structure for \texttt{4jpy} exhibits a chemically plausible backbone conformation that effectively accommodates the native ligand from the PDBbind dataset under rigid-body docking conditions. The resulting docking poses reveal appropriate spatial accommodation and shape complementarity between the ligand and the predicted binding pocket. According to the docking evaluation in Table~\ref{tab:qdock_4jpy}, the predicted docking poses from \textbf{QDockBank} achieve an average binding affinity of \textbf{-4.3 kcal/mol}, which is lower than that of AlphaFold3 (\textbf{-3.9 kcal/mol}). This indicates stronger ligand-binding potential for the QDockBank-predicted fragment and better alignment with its biological functionality. Moreover, the RMSD lower bound (l.b.) and upper bound (u.b.) reflect the structural variability across possible docking poses in a single docking run. Smaller RMSD l.b. and u.b. values suggest greater structural stability and superior prediction quality. For QDockBank, both average RMSD l.b. and u.b. remain below \textbf{2.0~\AA}, indicating stable binding conformations. In contrast, AlphaFold3 predictions for the same fragment exhibit larger variation, with RMSD ranges spanning from \textbf{1.9~\AA} to \textbf{3.2~\AA}, and both the lower and upper bounds being consistently higher than those from QDockBank. This demonstrates that the AF3-predicted structure is less stable and less reliable in docking scenarios.

\begin{table}[htbp]
\centering
\caption{Average docking metrics for QDockBank vs. AlphaFold3 on \texttt{4jpy}}
\label{tab:qdock_4jpy}
\resizebox{\linewidth}{!}{
\begin{tabular}{l|c|c}
\toprule
Metric & QDockBank & AlphaFold3 \\
\midrule
Affinity (kcal/mol)(Low is better) & -4.3 & -3.9 \\
RMSD l.b. (\AA)(Low is better)     & 1.4  & 2.0 \\
RMSD u.b. (\AA)(Low is better)     & 1.9  & 3.2 \\
\bottomrule
\end{tabular}
}
\end{table}

\subsection{RMSD-Based Structural Evaluation and Visualization}

\begin{figure}[htbp]
    \centering
    \begin{subfigure}[t]{0.45\linewidth}
        \centering
        \includegraphics[trim=105 20 105 30, clip, width=\linewidth]{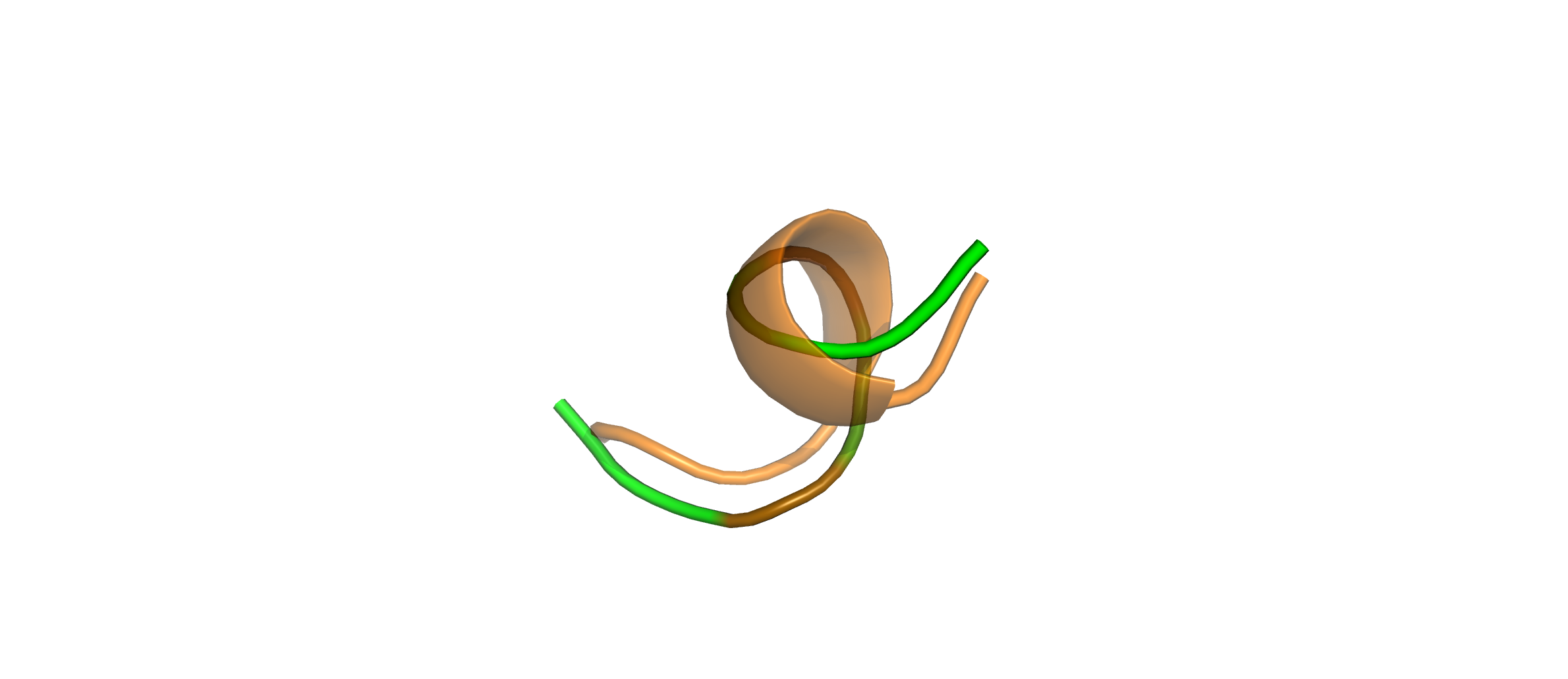}
        \caption{\textit{QDockBank} Prediction}
    \end{subfigure}
    \hfill
    \begin{subfigure}[t]{0.45\linewidth}
        \centering
        \includegraphics[trim=105 20 105 30, clip, width=\linewidth]{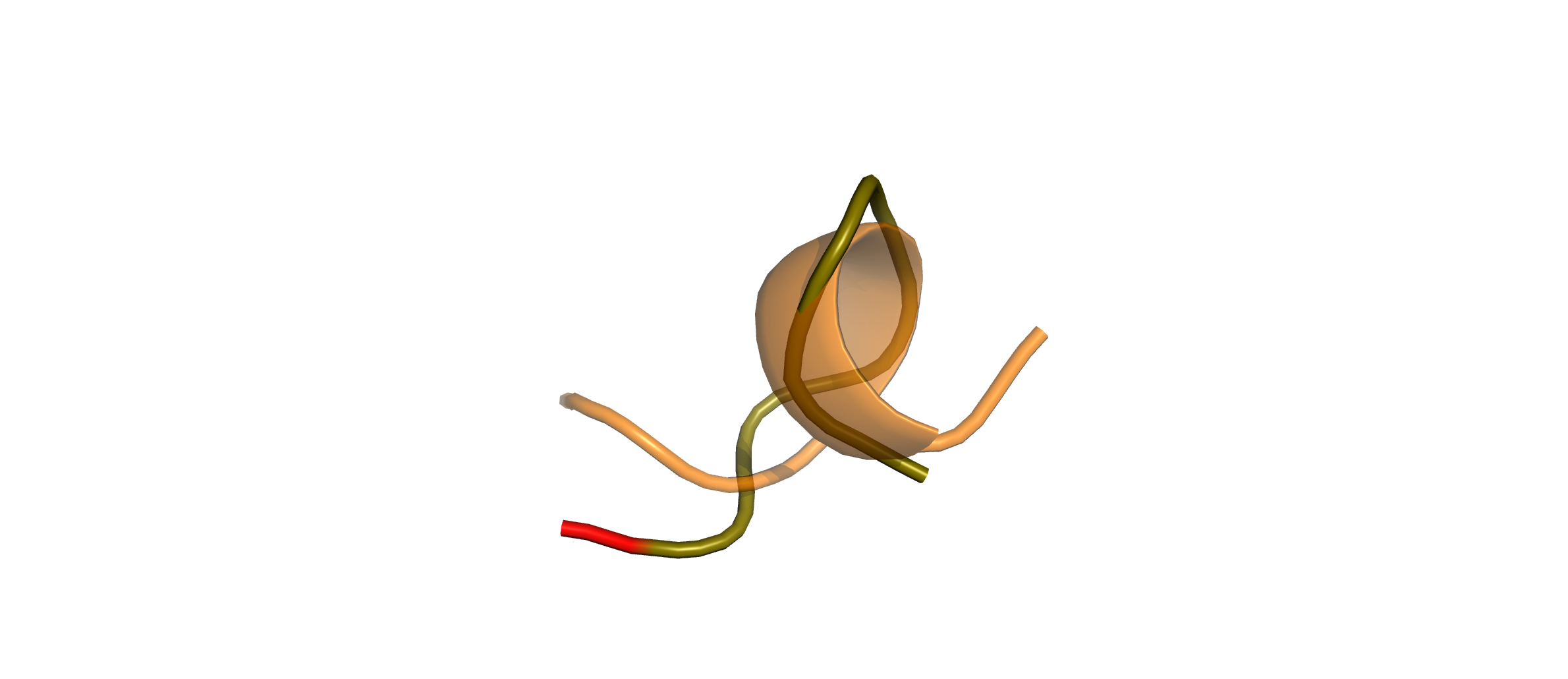}
        \caption{AlphaFold3 Prediction}
    \end{subfigure}
    
    \caption{RMSD-based structural comparison of the predicted fragment for PDB entry \texttt{2qbs}. The experimental structure (orange) is compared with \textit{QDockBank} (left) and AlphaFold3 (right) predictions. Regions with close alignment are shown in green, while structural deviations are highlighted in red. QDockBank exhibits higher structural accuracy than AlphaFold3.}
    \label{fig7}
\end{figure}

In addition to docking performance, another critical application of \textit{QDockBank} is the evaluation of structural prediction accuracy. Each predicted fragment is paired with its experimentally determined counterpart obtained via X-ray crystallography, enabling researchers to quantify structural deviations using root-mean-square deviation (RMSD)~\cite{cohen1980prediction}. All PDB files in \textit{QDockBank} adhere strictly to the PDB format specification, ensuring compatibility with external databases such as RCSB PDB and PDBbind. RMSD can be computed over C$\alpha$ atoms, backbone atoms, or all atoms, depending on the desired resolution and specificity. Predicted structures can also be visualized directly using standard molecular visualization tools such as PyMOL~\cite{delano2002pymol}, Chimera~\cite{pettersen2004ucsf}, or VMD~\cite{humphrey1996vmd}, supporting both automated validation and manual inspection in modeling workflows and publications. As illustrated in Figure~\ref{fig7}, we present an RMSD-based structural comparison involving a fragment from PDB entry \texttt{2qbs}. In this figure, the experimentally determined structure is shown in orange, alongside predicted structures from \textit{QDockBank} and AlphaFold3. Green regions indicate local agreement with the reference structure, while red highlights structural deviations between the predicted and experimental conformations.

The fragment corresponds to residues 217–224 of chain A in PDB entry \texttt{2qbs}, encompassing a canonical alpha-helical segment. When aligning the predicted structures from QDock and AlphaFold3 (AF3) with the experimentally determined conformation, the QDock prediction closely follows the experimental structure from residues 217 to 220. Within the helical region spanning residues 221–223, the QDock model maintains accurate structural alignment, supporting the effectiveness of sequence-driven quantum modeling. In contrast, the AF3-predicted fragment exhibits substantial deviations between residues 217 and 222. In particular, the helical segment spanning residues 221–223 is misaligned, with residue 222 disrupting the helical conformation and diminishing the structural plausibility of the entire segment. Quantitatively, the final RMSD of the QDock prediction is 2.428~\AA, while that of the AF3 prediction is 4.234~\AA—representing nearly a twofold compared to the QDock model.

\section{Conclusion}

\textit{QDockBank} provides the first large-scale dataset of protein fragments generated entirely using utility-level quantum processors, offering biologically relevant, docking-ready models for structure-based drug discovery. By integrating quantum modeling, docking evaluation, and RMSD-based assessment, it enables systematic benchmarking of quantum-assisted structure prediction. With broad protein type coverage and complete amino acid interaction profiles, \textit{QDockBank} fills a critical gap and provides a reproducible foundation for advancing quantum applications in computational biology.

\clearpage

\bibliographystyle{unsrt}
\bibliography{reference}

\begin{thebibliography}{10}

\bibitem{liang1998anatomy}
Jie Liang, Clare Woodward, and Herbert Edelsbrunner.
\newblock Anatomy of protein pockets and cavities: measurement of binding site geometry and implications for ligand design.
\newblock {\em Protein science}, 7(9):1884--1897, 1998.

\bibitem{abramson2024accurate}
Josh Abramson, Jonas Adler, Jack Dunger, Richard Evans, Tim Green, Alexander Pritzel, Olaf Ronneberger, Lindsay Willmore, Andrew~J Ballard, Joshua Bambrick, et~al.
\newblock Accurate structure prediction of biomolecular interactions with alphafold 3.
\newblock {\em Nature}, 630(8016):493--500, 2024.

\bibitem{jumper2021highly}
John Jumper, Richard Evans, Alexander Pritzel, Tim Green, Michael Figurnov, Olaf Ronneberger, Kathryn Tunyasuvunakool, Russ Bates, Augustin {\v{Z}}{\'\i}dek, Anna Potapenko, et~al.
\newblock Highly accurate protein structure prediction with alphafold.
\newblock {\em nature}, 596(7873):583--589, 2021.

\bibitem{evans2021protein}
Richard Evans, Michael O’Neill, Alexander Pritzel, Natasha Antropova, Andrew Senior, Tim Green, Augustin {\v{Z}}{\'\i}dek, Russ Bates, Sam Blackwell, Jason Yim, et~al.
\newblock Protein complex prediction with alphafold-multimer.
\newblock {\em biorxiv}, pages 2021--10, 2021.

\bibitem{wang2024efficient}
Youle Wang and Xiangzhen Zhou.
\newblock Efficient quantum algorithm for lattice protein folding.
\newblock {\em Quantum Science and Technology}, 10(1):015056, 2024.

\bibitem{shoichet1991protein}
Brian~K Shoichet and Irwin~D Kuntz.
\newblock Protein docking and complementarity.
\newblock {\em Journal of molecular biology}, 221(1):327--346, 1991.

\bibitem{robert2021resource}
Anton Robert, Panagiotis~Kl Barkoutsos, Stefan Woerner, and Ivano Tavernelli.
\newblock Resource-efficient quantum algorithm for protein folding.
\newblock {\em npj Quantum Information}, 7(1):38, 2021.

\bibitem{doga2024perspective}
Hakan Doga, Bryan Raubenolt, Fabio Cumbo, Jayadev Joshi, Frank~P DiFilippo, Jun Qin, Daniel Blankenberg, and Omar Shehab.
\newblock A perspective on protein structure prediction using quantum computers.
\newblock {\em Journal of Chemical Theory and Computation}, 20(9):3359--3378, 2024.

\bibitem{berman2007worldwide}
Helen Berman, Kim Henrick, Haruki Nakamura, and John~L Markley.
\newblock The worldwide protein data bank (wwpdb): ensuring a single, uniform archive of pdb data.
\newblock {\em Nucleic acids research}, 35(suppl\_1):D301--D303, 2007.

\bibitem{sussman1998protein}
Joel~L Sussman, Dawei Lin, Jiansheng Jiang, Nancy~O Manning, Jaime Prilusky, Otto Ritter, and Enrique~E Abola.
\newblock Protein data bank (pdb): database of three-dimensional structural information of biological macromolecules.
\newblock {\em Biological Crystallography}, 54(6):1078--1084, 1998.

\bibitem{burley2017protein}
Stephen~K Burley, Helen~M Berman, Gerard~J Kleywegt, John~L Markley, Haruki Nakamura, and Sameer Velankar.
\newblock Protein data bank (pdb): the single global macromolecular structure archive.
\newblock {\em Protein crystallography: methods and protocols}, pages 627--641, 2017.

\bibitem{berman2002protein}
Helen~M Berman, Tammy Battistuz, Talapady~N Bhat, Wolfgang~F Bluhm, Philip~E Bourne, Kyle Burkhardt, Zukang Feng, Gary~L Gilliland, Lisa Iype, Shri Jain, et~al.
\newblock The protein data bank.
\newblock {\em Biological Crystallography}, 58(6):899--907, 2002.

\bibitem{orengo1999cath}
Christine~A. Orengo, Frances M.~G. Pearl, James~E. Bray, Annabel~E. Todd, AC~Martin, L~Lo~Conte, and Janet~M. Thornton.
\newblock The cath database provides insights into protein structure/function relationships.
\newblock {\em Nucleic acids research}, 27(1):275--279, 1999.

\bibitem{pearl2003cath}
Frances~MG Pearl, CF~Bennett, James~E Bray, Andrew~P Harrison, Nigel Martin, A~Shepherd, Ian Sillitoe, J~Thornton, and Christine~A Orengo.
\newblock The cath database: an extended protein family resource for structural and functional genomics.
\newblock {\em Nucleic acids research}, 31(1):452--455, 2003.

\bibitem{murzin1995scop}
Alexey~G Murzin, Steven~E Brenner, Tim Hubbard, and Cyrus Chothia.
\newblock Scop: a structural classification of proteins database for the investigation of sequences and structures.
\newblock {\em Journal of molecular biology}, 247(4):536--540, 1995.

\bibitem{varadi2022alphafold}
Mihaly Varadi, Stephen Anyango, Mandar Deshpande, Sreenath Nair, Cindy Natassia, Galabina Yordanova, David Yuan, Oana Stroe, Gemma Wood, Agata Laydon, et~al.
\newblock Alphafold protein structure database: massively expanding the structural coverage of protein-sequence space with high-accuracy models.
\newblock {\em Nucleic acids research}, 50(D1):D439--D444, 2022.

\bibitem{Mihaly2024alphafold}
Mihaly Varadi, Damian Bertoni, Paulyna Magana, Urmila Paramval, Ivanna Pidruchna, Malarvizhi Radhakrishnan, Maxim Tsenkov, Sreenath Nair, Milot Mirdita, Jingi Yeo, et~al.
\newblock Alphafold protein structure database in 2024: providing structure coverage for over 214 million protein sequences.
\newblock {\em Nucleic acids research}, 52(D1):D368--D375, 2024.

\bibitem{holcomb2023evaluation}
Matthew Holcomb, Ya-Ting Chang, David~S Goodsell, and Stefano Forli.
\newblock Evaluation of alphafold2 structures as docking targets.
\newblock {\em Protein Science}, 32(1):e4530, 2023.

\bibitem{gfeller2012swisssidechain}
David Gfeller, Olivier Michielin, and Vincent Zoete.
\newblock Swisssidechain: a molecular and structural database of non-natural sidechains.
\newblock {\em Nucleic acids research}, 41(D1):D327--D332, 2012.

\bibitem{wang2005pdbbind}
Renxiao Wang, Xueliang Fang, Yipin Lu, Chao-Yie Yang, and Shaomeng Wang.
\newblock The pdbbind database: methodologies and updates.
\newblock {\em Journal of medicinal chemistry}, 48(12):4111--4119, 2005.

\bibitem{liu2015pdb}
Zhihai Liu, Yan Li, Li~Han, Jie Li, Jie Liu, Zhixiong Zhao, Wei Nie, Yuchen Liu, and Renxiao Wang.
\newblock Pdb-wide collection of binding data: current status of the pdbbind database.
\newblock {\em Bioinformatics}, 31(3):405--412, 2015.

\bibitem{trott2010autodock}
Oleg Trott and Arthur~J Olson.
\newblock Autodock vina: improving the speed and accuracy of docking with a new scoring function, efficient optimization, and multithreading.
\newblock {\em Journal of computational chemistry}, 31(2):455--461, 2010.

\bibitem{yue2024integration}
Yang Yue, Shu Li, Yihua Cheng, Lie Wang, Tingjun Hou, Zexuan Zhu, and Shan He.
\newblock Integration of molecular coarse-grained model into geometric representation learning framework for protein-protein complex property prediction.
\newblock {\em Nature Communications}, 15(1):9629, 2024.

\bibitem{tilly2022variational}
Jules Tilly, Hongxiang Chen, Shuxiang Cao, Dario Picozzi, Kanav Setia, Ying Li, Edward Grant, Leonard Wossnig, Ivan Rungger, George~H Booth, et~al.
\newblock The variational quantum eigensolver: a review of methods and best practices.
\newblock {\em Physics Reports}, 986:1--128, 2022.

\bibitem{joshi2021evaluating}
Nisheeth Joshi, Pragya Katyayan, and Syed~Afroz Ahmed.
\newblock Evaluating the performance of some local optimizers for variational quantum classifiers.
\newblock In {\em Journal of Physics: Conference Series}, volume 1817, page 012015. IOP Publishing, 2021.

\bibitem{o2011open}
Noel~M O'Boyle, Michael Banck, Craig~A James, Chris Morley, Tim Vandermeersch, and Geoffrey~R Hutchison.
\newblock Open babel: An open chemical toolbox.
\newblock {\em Journal of cheminformatics}, 3:1--14, 2011.

\bibitem{chow2021ibm}
Jerry Chow, Oliver Dial, and Jay Gambetta.
\newblock Ibm quantum breaks the 100-qubit processor barrier.
\newblock {\em IBM Research Blog}, 2, 2021.

\bibitem{kim2023evidence}
Youngseok Kim, Andrew Eddins, Sajant Anand, Ken~Xuan Wei, Ewout Van Den~Berg, Sami Rosenblatt, Hasan Nayfeh, Yantao Wu, Michael Zaletel, Kristan Temme, et~al.
\newblock Evidence for the utility of quantum computing before fault tolerance.
\newblock {\em Nature}, 618(7965):500--505, 2023.

\bibitem{siddiqi2021engineering}
Irfan Siddiqi.
\newblock Engineering high-coherence superconducting qubits.
\newblock {\em Nature Reviews Materials}, 6(10):875--891, 2021.

\bibitem{li2023error}
Zhiyuan Li, Pei Liu, Peng Zhao, Zhenyu Mi, Huikai Xu, Xuehui Liang, Tang Su, Weijie Sun, Guangming Xue, Jing-Ning Zhang, et~al.
\newblock Error per single-qubit gate below 10- 4 in a superconducting qubit.
\newblock {\em npj Quantum Information}, 9(1):111, 2023.

\bibitem{zhou2020quantum}
Leo Zhou, Sheng-Tao Wang, Soonwon Choi, Hannes Pichler, and Mikhail~D Lukin.
\newblock Quantum approximate optimization algorithm: Performance, mechanism, and implementation on near-term devices.
\newblock {\em Physical Review X}, 10(2):021067, 2020.

\bibitem{li2019tackling}
Gushu Li, Yufei Ding, and Yuan Xie.
\newblock Tackling the qubit mapping problem for nisq-era quantum devices.
\newblock In {\em Proceedings of the twenty-fourth international conference on architectural support for programming languages and operating systems}, pages 1001--1014, 2019.

\bibitem{li2020qubit}
Sanjiang Li, Xiangzhen Zhou, and Yuan Feng.
\newblock Qubit mapping based on subgraph isomorphism and filtered depth-limited search.
\newblock {\em IEEE Transactions on Computers}, 70(11):1777--1788, 2020.

\bibitem{wang2004pdbbind}
Renxiao Wang, Xueliang Fang, Yipin Lu, and Shaomeng Wang.
\newblock The pdbbind database: Collection of binding affinities for protein- ligand complexes with known three-dimensional structures.
\newblock {\em Journal of medicinal chemistry}, 47(12):2977--2980, 2004.

\bibitem{cock2009biopython}
Peter~JA Cock, Tiago Antao, Jeffrey~T Chang, Brad~A Chapman, Cymon~J Cox, Andrew Dalke, Iddo Friedberg, Thomas Hamelryck, Frank Kauff, Bartek Wilczynski, et~al.
\newblock Biopython: freely available python tools for computational molecular biology and bioinformatics.
\newblock {\em Bioinformatics}, 25(11):1422, 2009.

\bibitem{chapman2000biopython}
Brad Chapman and Jeffrey Chang.
\newblock Biopython: Python tools for computational biology.
\newblock {\em ACM Sigbio Newsletter}, 20(2):15--19, 2000.

\bibitem{kuroda2016pushing}
Daisuke Kuroda and Jeffrey~J Gray.
\newblock Pushing the backbone in protein-protein docking.
\newblock {\em Structure}, 24(10):1821--1829, 2016.

\bibitem{gilson2007calculation}
Michael~K Gilson and Huan-Xiang Zhou.
\newblock Calculation of protein-ligand binding affinities.
\newblock {\em Annu. Rev. Biophys. Biomol. Struct.}, 36(1):21--42, 2007.

\bibitem{mirdita2022colabfold}
Milot Mirdita, Konstantin Sch{\"u}tze, Yoshitaka Moriwaki, Lim Heo, Sergey Ovchinnikov, and Martin Steinegger.
\newblock Colabfold: making protein folding accessible to all.
\newblock {\em Nature methods}, 19(6):679--682, 2022.

\bibitem{miyazawa1985estimation}
Sanzo Miyazawa and Robert~L Jernigan.
\newblock Estimation of effective interresidue contact energies from protein crystal structures: quasi-chemical approximation.
\newblock {\em Macromolecules}, 18(3):534--552, 1985.

\bibitem{cohen1980prediction}
Fred~E Cohen and Michael~JE Sternberg.
\newblock On the prediction of protein structure: the significance of the root-mean-square deviation.
\newblock {\em Journal of molecular biology}, 138(2):321--333, 1980.

\bibitem{delano2002pymol}
Warren~L DeLano et~al.
\newblock Pymol: An open-source molecular graphics tool.
\newblock {\em CCP4 Newsl. Protein Crystallogr}, 40(1):82--92, 2002.

\bibitem{pettersen2004ucsf}
Eric~F Pettersen, Thomas~D Goddard, Conrad~C Huang, Gregory~S Couch, Daniel~M Greenblatt, Elaine~C Meng, and Thomas~E Ferrin.
\newblock Ucsf chimera—a visualization system for exploratory research and analysis.
\newblock {\em Journal of computational chemistry}, 25(13):1605--1612, 2004.

\bibitem{humphrey1996vmd}
William Humphrey, Andrew Dalke, and Klaus Schulten.
\newblock Vmd: visual molecular dynamics.
\newblock {\em Journal of molecular graphics}, 14(1):33--38, 1996.

\end{thebibliography}




\end{document}